\documentclass[11pt,a4paper]{article}
\usepackage{epsf}

\usepackage{epsfig,graphics}
\usepackage{graphicx}
\usepackage{epsfig}
\usepackage{subfigure}
\usepackage{tabularx} 
\usepackage{rotate}	
\usepackage{slashed}
\usepackage{color}

\usepackage{color}
\usepackage[usenames,dvipsnames,svgnames,table]{xcolor}
\usepackage[colorlinks=true,
            linkcolor=blue,
            urlcolor=gray,
            citecolor=green]{hyperref}

\usepackage{amsmath}
\usepackage{amsfonts}
\usepackage{amssymb}
\usepackage{graphicx}
\usepackage{cite}

\usepackage[utf8]{inputenc}

\def\beq{\begin{equation}}
\def\eeq{\end{equation}}
\def\beqa{\begin{eqnarray}}
\def\eeqa{\end{eqnarray}}

\begin{document}

\makeatletter
\@addtoreset{equation}{section}
\makeatother
\renewcommand{\theequation}{\thesection.\arabic{equation}}
\pagestyle{empty}
\rightline{MPP-2016-319}
 \rightline{}
\vspace{1.2cm}
\begin{center}
\LARGE{\bf Backreaction Issues in Axion Monodromy and Minkowski 4-forms \\[12mm]}
\large{ Irene Valenzuela\\[4mm]}
\footnotesize{
 Max-Planck-Institut fur Physik,
Fohringer Ring 6, 80805 Munich, Germany\\
Institute for Theoretical Physics and
Center for Extreme Matter and Emergent Phenomena,\\
Utrecht University, Leuvenlaan 4, 3584 CE Utrecht, The Netherlands}

\vspace*{1cm}

\small{\bf Abstract} \\[5mm]
\end{center}
\begin{center}
\begin{minipage}[h]{16.0cm}
We clarify the differences between the usual Kaloper-Sorbo description of axion monodromy and the effective axionic potential in terms of Minkowski 4-forms derived in string compactifications. The fact that the metric of the 3-form fields coming from string theory is  field dependent (unlike in Kaloper-Sorbo) leads to the backreaction issues recently studied in axion monodromy models within string theory. We reanalyse these problems in terms of the 4-forms focusing on the case in which the non-periodic scalars backreact on the Kahler metric of the inflaton reducing the physical field range. In the closed string sector of Type II Calabi-Yau compactifications with fluxes the metric becomes field dependent precisely when $\Delta\phi\sim M_p$, independently of the choice of fluxes. We propose, however, some counter-examples to this universal behaviour by including open string fields.
\end{minipage}
\end{center}
\newpage
\setcounter{page}{1}
\pagestyle{plain}
\renewcommand{\thefootnote}{\arabic{footnote}}
\setcounter{footnote}{0}
\vspace*{1cm}

\tableofcontents

\section{Introduction}

Axion monodromy is one of the most promising proposals to achieve transplanckian field ranges in string theory \cite{Silverstein:2008sg,McAllister:2008hb} (see also \cite{Flauger:2009ab,Berg:2009tg,Pajer:2013fsa,Gur-Ari:2013sba,Palti:2014kza,Marchesano:2014mla,Blumenhagen:2014gta,Hebecker:2014eua,Ibanez:2014kia,Arends:2014qca,McAllister:2014mpa,Franco:2014hsa,Blumenhagen:2014nba,Hayashi:2014aua,Hebecker:2014kva,Ibanez:2014swa,Garcia-Etxebarria:2014wla,Blumenhagen:2015kja,Retolaza:2015sta,Escobar:2015fda,Blumenhagen:2015xpa,Escobar:2015ckf,Hebecker:2015tzo,Landete:2016cix}). It is based on having an axion with a multi-branched/multi-valued potential in such a way that the effective theory preserves the discrete shift symmetry of the axion if combined with appropriate shifts of the parameters of the potential. This multibranched structure unfolds the compact moduli space of the axion, allowing (a priori) for a transplanckian field range even if the fundamental periodicity (given by the axion decay constant) is subplanckian. Axion monodromy models can be described in terms of an axion coupled to a Minkowski 3-form field as follows \cite{Kaloper:2008fb,Kaloper:2011jz,Marchesano:2014mla,Dudas:2014pva}, 
\beq
\mathcal{L}=-f^2(d\phi)^2- F_4\wedge *F_4+2F_4\phi
\label{KS}
\eeq
where $f$ is the decay constant of the axion. 
Upon integrating out the Minkowski 3-form (which has no propagating degrees of freedom in four dimensions) a scalar potential for the axion is generated with the aforementioned multi-branched structure,
\beq
*F_4=f_0+m\phi\rightarrow V=(f_0+m\phi)^2
\eeq
Here $f_0$ is an integration constant corresponding to a possible constant value of the 4-form field strength in the vacuum. The above scalar potential is indeed invariant under the combined global shift 
\beq
f_0\rightarrow f_0 + c\ ,\quad \phi\rightarrow \phi -c/m
\eeq
which relates different branches labelled by $f_0$. When $c/m=2\pi f$ the above transformation relates gauge equivalent branches identified by the discrete periodicity of the axion. In the presence of membranes electrically charged under the 3-form gauge field, the constant $c$ is quantized in units of the 3-form gauge coupling $\Lambda_k$, implying in turn the following consistency relation\footnote{If $n\neq 1$, the original continuous global symmetry of the axion is broken to a discrete global symmetry $Z_n$ by the coupling to the 3-form field. Therefore, the discrete periodicity of the axion is always preserved at the level of the theory.} $2\pi f=  n\Lambda_k^2/m$. This is what is commonly known as the Kaloper-Sorbo model since they were the first ones in using such a description in terms of a 3-form field for inflation \cite{Kaloper:2008fb,Kaloper:2011jz,Kaloper:2014zba,Kaloper:2016fbr} (see, however, also the earlier work of Dvali \cite{Dvali:2005an,Dvali:2005zk}).

It is important to remark that the discrete shift symmetry of the axion is preserved by the system and only broken spontaneously (not explicitly) upon selecting a vacuum.
At classical level, $f_0$ is fixed, which selects a concrete branch and consequently a particular vacuum. One can then classically roll down a single branch for a field range much bigger than $2\pi f$. However, at quantum level the different branches are dynamically connected to each other since one can induce quantum tunneling transitions between them by nucleation of membranes \cite{Coleman:1980aw,Bousso:2000xa,Brown:1988kg,Duncan:1989ug,Feng:2000if}. By crossing a membrane of charge $k$ one shifts he value of $f_0$ by $k$ units. This tunneling between branches reduces the effective field range of the axion ruling out parametrically large displacements. The concrete bound on the field range will depend on the specific value of the tension of the membrane, which can be estimated for instance by using the Weak Gravity Conjecture \cite{ArkaniHamed:2006dz} or specific UV completions in string theory. In \cite{Ibanez:2015fcv,Hebecker:2015zss,Brown:2016nqt} it was shown that generically the tunneling rate is highly suppressed and these 'jumps' do not generate a problem for large field inflation.

On the other hand, the gauge invariance of the 3-form highly constraints higher order corrections to the above Lagrangian \cite{Kaloper:2011jz}. Only corrections depending on the gauge invariant field strength are allowed, which implies that, upon integrating out the 3-form, the corrections to the scalar potential goes as powers of the potential itself,
\beq
\delta V = \left(\frac{V_0}{M^4_p}\right)^n 
\label{corrections}
\eeq
Therefore even if $\phi$ takes transplanckian values, as long as the potential energy remains subplanckian, the corrections remain under control. This is consistent with the fact that the discrete shift symmetry of the axion must be gauged in a consistent theory of quantum gravity and therefore cannot be explicitly, but only spontaneously, broken. Since the source of spontaneous breaking can always be parametrized in four dimensions by coupling the axion to an effective 4-form field strength, all corrections have to appear as functions of the field strength itself, which on shell is dual to the shift invariant function $\rho=m\phi+f_0$.  In the case of a single axion and a single 4-form this leads to \eqref{corrections}. We will see that in the presence of multiple 4-forms (multiple sources of spontaneous breaking) the corrections will be proportional to the different parts of the potential induced by each 4-form, so the corrections will take a more elaborated form than in \eqref{corrections}. But again, as long as the energy density is subplanckian, they will be subleading. No need to mention that one can also have shift-symmetric non-perturbative corrections like $\sim cos(\phi)$. But since they give rise to periodic bounded potentials the effective theory remains under control. In any case, corrections going as powers of $\phi/M_p$ that grow parametrically with the field overrunning the tree-level potential are forbidden because of the gauge symmetries of the system.

This outstanding protection under higher order corrections makes axion monodromy a promising proposal to get transplanckian field ranges in a controlled way. Besides, as explained above, constraints coming from the Weak Gravity Conjecture are much weaker than in models based on natural inflation with one or multiple axions \cite{delaFuente:2014aca,Rudelius:2014wla,Rudelius:2015xta,Montero:2015ofa,Brown:2015iha,Bachlechner:2015qja,Hebecker:2015rya,Brown:2015lia,Junghans:2015hba,Palti:2015xra,Heidenreich:2015nta,Kooner:2015rza,Heidenreich:2015wga,Heidenreich:2016aqi,Montero:2016tif,Hebecker:2016dsw,Saraswat:2016eaz}. However, all attempts to realise axion monodromy in string theory are not free of problems and technical difficulties \cite{McAllister:2014mpa,Hebecker:2014kva,Blumenhagen:2014nba,Buchmuller:2015oma,Andriot:2015aza}. In some cases, the backreaction of the rest of the moduli of the compactification can have non-negligible effects and reduce drastically the effective field range \cite{palti}. In this sense, the presence of a Kaloper-Sorbo coupling is not enough to guarantee a transplanckian field range. What is then missing in our above description of axion monodromy in terms of an axion coupled to a Minkowski 3-form? 

In this paper we will explain the differences between the Kaloper-Sorbo model above and the effective theories that one actually obtain from string compactifications, with the aim of clarifying the advantages and drawbacks of the construction. Remarkably, the effective axionic potential coming from string theory can always be rewritten as a generalization of Kaloper-Sorbo with non-canonical field-dependent metrics \cite{4forms,Carta:2016ynn}. It is this field-dependence on the kinetic metric of the 3-form field, not present in the original Kaloper-Sorbo model but characteristic of any supergravity generalization of \eqref{KS}, what generates in the end the backreaction issues found in particular string realisations of axion monodromy. Here we will reanalysed these problems motivated by the question whether these difficulties are simply technical limitations of particular models or instead a hint of an underlying obstruction of having a transplanckian field range in a consistent theory of quantum gravity. The reformulation of the backreaction issues in the dual picture in terms of 3-form fields allows for a more model-independent analysis of these difficulties, hopefully shedding some light on this topic. 

This paper is organized as follows. In the next section we will explain the new ingredients appearing when constructing a supergravity description of the original bosonic model of Kaloper-Sorbo, which will also appear in any $N=1$ string compactification and can lead to the backreaction issues discussed above. In section 3 we will review how the backreaction coming from other non-periodic fields can reduce the effective field range by redefining the proper field distance of the inflaton, and explain how this effect is encoded on having field-dependent kinetic metrics for the 3-form gauge fields. Whether this backreaction rules out transplanckian or only infinite (parametrically large) excursions depends on the concrete form of the kinetic metric, which will be related to the Kahler metric of the scalar manifold in an $N=1$ compactification. Therefore, in section 4, we will analyse the effective theories derived in flux compactifications of Type IIA(B). We reproduce the results of \cite{palti} and find that the proper field distance for closed string axions grow at best logarithmically at large field and, remarkably, this logarithmic behaviour appears shortly after crossing $\phi \sim M_p$. However, we argue in section 5 that the presence of open string fields might provide a counter-example to this universal behaviour since, a priori, one can delay the logarithmic behaviour far away in field distance by tuning the fluxes. We leave section 6 for conclusions.

\section{Generalization of Kaloper-Sorbo structure in string theory}

As we reviewed in the Introduction, one can provide a mass term to an axion without breaking explicitly the shift symmetry and without adding new degrees of freedom, simply by coupling the axion to a 4-form field strength living in the four dimensional space-time \cite{Quevedo:1996uu,Dvali:2005an}. In fact, if the discrete shift symmetry of the axion is preserved, one can always write such a dual description in terms of a 3-form gauge field. Interestingly, this is indeed the way in which axions get a perturbative potential in string compactifications\footnote{Here we are writing only the scalar potential at perturbative level, but there can also be non-perturbative corrections leading to additional periodic (shift-symmetric) potentials for the axions. Interestingly, these terms can also be written in terms of effective 3-form fields, corresponding for instance to a composite Chern-Simons 3-form in the case of non-perturbative gauge dynamics \cite{Dvali:2005zk,Dvali:2005an}. The formulation in terms of 3-form fields for the case of non-gauge D-brane instanton effects has been performed in \cite{Garcia-Valdecasas:2016voz}.}.
In \cite{4forms} it was shown that the four dimensional effective theory coming from compactifications of Type IIA(B) on a Calabi-Yau 3-fold with orientifolds and fluxes can be written in the form
\beq
V=-Z_{ab}(s^i)F_4^a\wedge *F_4^b +2F_4^a\rho_a(\phi^i)+V_{loc}(s^i)
\label{VII}
\eeq
where all the dependence of the scalar potential on the axions $\phi^i$ comes from couplings to 3-form gauge fields. In the above potential, $s^i$ stand for the saxions (non-periodic scalars of the compactification), $i,j$ runs over all the moduli and $a,b$ over all the 3-form fields. In flux compactifications, the 3-forms are not effective or composite fields, but have a clear microscopic interpretation: they come from dimensional reduction of the higher RR and NSNS p-form fields. Therefore they are dual to the internal fluxes of the compactification, which are known to induce an F-term potential for the axions. Even if the axions only appear through couplings to 3-forms fields, the scalar potential for the saxions can receive other contributions from other elements of the compactification, which we include in $V_{loc}(s^i)$. Upon integrating out the 3-form fields, one gets
\beq
V=Z_{ab}(s^i)^{-1}\rho_a(\phi^i)\rho_b(\phi^i)+V_{loc}(s^i)
\eeq
which can be rearranged to recover the usual $N=1$ Cremmer et al. scalar potential of Type IIA(B). In the following we will highlight the differences between this effective theory and non-supersymmetric Kaloper-Sorbo model \eqref{KS}.
\begin{itemize}
\item Non-linear couplings $\rightarrow$ Generic scalar potentials

In Kaloper-Sorbo, the axion couples linearly to $F_4$ which induces a mass term for the axion in the effective theory. However, one can have more general couplings which induce, not only mass terms, but also cubic or higher couplings in the effective theory. In a string compactification, each 3-form couples to a function $\rho_a(\phi^i)$ which only depends on the axions (and not the saxions) and topological data of the compactification. This function has to be invariant under the discrete shifts of the axions (corresponding to large gauge transformations in the higher dimensional theory), implying that the parameters inside $\rho_a(\phi^i)$ have to transform accordingly to reabsorb these shifts. In flux compactifications, these parameters correspond to the internal flux quanta which indeed enjoy the appropriate shift transformations to keep each $\rho_a(\phi^i)$ invariant.

\item Multiple 3-forms $\rightarrow$ Higher order corrections

The presence of multiple 3-forms implies that the higher order corrections will appear as products or combinations of the different 4-form field strengths. Therefore the corrections do not necessarily appear as powers of the potential itself, but as powers of combinations of the different $\rho(\phi)$, which are shift invariant by themselves. Therefore the corrections are under control if all the functions $\rho_a(\phi)$ remain subplanckian. This still protects large field inflation over tranplanckian field ranges of the inflaton, but makes the computation more technically involved, because a priori one needs then to know all the 3-forms to whom the inflaton couples, in order to have control over all the $\rho_a(\phi)$. However, as long as the different parts of the potential remain subplanckian, the corrections will always be subleading. They can though have important implications for inflation and modify the scalar potential, leading for instance to interesting flattening effects (see e.g. \cite{Bielleman:2016grv}).

\item Non-trivial field-dependent metrics  $\rightarrow$ Backreaction

The main difference with respect to the non-supersymmetric Kaloper-Sorbo model  comes from the presence of other non-periodic scalars, which we call saxions since they usually combine with the axions to fill $N=1$ chiral multiplets. As usual in supergravity, the kinetic metric of the 3-form fields is field-dependent and only in the case in which there is a large mass hierarchy between the inflaton and the saxions, one can treat these fields as fixed parameters and recover approximately the Kaloper-Sorbo model of \eqref{KS}. However, it is precisely the backreaction of the saxions what has been proved to reduce drastically the effective field range in some string axion monodromy models. Therefore, we should reanalyse the problem without assuming the saxions as fixed parameters, but as fields that are also stabilized due to the presence of the 3-form fields. This is the task for the next section.

\end{itemize}

\section{Backreaction effects in terms of 3-form fields}

In order to have a well-defined and controlled effective theory of large field inflation in the IR, one needs to either fine-tune an infinite number of parameters corresponding to the infinite tower of non-renormalizable operators or invoke the presence of some UV symmetry that forbids these operators and protects the potential from large transplanckian excursions. Periodic real scalars (axions) with their shift symmetries are then promising candidates for large field inflation. A long-standing problem though in supersymmetric theories is how to stabilize the bosonic partner of the axion (the saxion) so that it does not spoil the dynamics of inflation. The typical solution is try to engineer a model in which the scale of moduli stabilization is higher than the inflationary scale, so the analysis can be divided in two steps. First, one stabilizes all the scalars except for the ones relevant for inflation, and secondly one studies the dynamics of the remaining fields assuming that the heavy moduli stay approximately fixed at their minimum values. However, this is not always a valid truncation of the theory. Sometimes the vacuum expectation values (vev) of the heavy moduli can depend on the inflaton vev in such a way that they lead to non-negligible modifications of the effective inflationary model when the inflaton is displaced away from its minimum. These backreaction effects have been recently studied in the context of string compactifications, see e.g. \cite{Buchmuller:2014vda,palti,Dudas:2015lga,Buchmuller:2015oma,McAllister:2016vzi}. The tricky part is to quantify these effects and extract general conclusions beyond the specific results of a particular model. Let us consider that the vev of a saxion depends on the value of the inflaton field as follows, 
\beq
\langle s\rangle = s_0 +\delta s(\phi)
\label{saxion1}
\eeq
This will lead inevitably to backreaction effects on the inflationary dynamics. For large values of the inflaton $\phi$ the displacement of $s$ from its minimum will be bigger, leading to a bigger modification of the effective theory which can affect both the potential and kinetic term of the inflaton. In general, one expects this displacement to depend on the mass ratio
\beq
\delta s(\phi) \sim f\left(\frac{m_\phi}{m_s} \phi^n\right)\label{saxion2}
\eeq
such that a big mass hierarchy suppresses the backreaction. Unfortunately, sometimes even a small hierarchy is difficult to engineer (see the no-go theorems postulated for the complex structure moduli space of Calabi-Yau 3-folds near the large complex structure point in \cite{Blumenhagen:2014nba,Hebecker:2014kva}).  In \cite{palti} it was claimed that in string theory one cannot delay indefinitely the backreaction effects by tuning the masses (which will be parametrized by fluxes in string Type II compactifications), but that the proper field distance of the inflaton will grow at best logarithmically with the inflaton vev as soon as field displacement becomes transplanckian. Let us review here the argument.
 The physical field distance is given by
\beq
\Delta\phi=\int K_{\phi,\bar \phi}^{1/2} d\phi
\eeq
where the metric  depends on the saxions of the compactification. If $K_{\phi,\bar \phi}\sim 1/s^2$ (like in typical string compactifications where $K=-log(s)$) and the vev of the saxion depends on the inflaton as in \eqref{saxion2} with $n>0$, then the kinetic metric will be inversely proportional to the value of the inflaton itself.  This leads to a redefinition of the canonical field reducing the effective field range. In particular if $\langle s\rangle = s_0 +\frac{m_\phi}{m_s} \phi$, then
 \beq
\Delta\phi=\int K_{\phi,\bar \phi}^{1/2} d\phi\sim \int \frac{1}{s(\phi)}d\phi\sim \int\frac{1}{s_0+\frac{m_\phi}{m_s}\phi}d\phi\sim log(s_0+\frac{m_\phi}{m_s}\phi)
\eeq
and the physical field distance scales at best logarithmically for large values of the field, ruling out parametrically large field ranges. This logarithmic behaviour\footnote{See also \cite{Blumenhagen:2015qda} for an analysis of the same type of canonical field redefinition in the context of Type IIB orientifolds with non-geometric fluxes. This logarithmic behaviour of the proper field distance implies that the effective potential becomes of Starobinsky-type at large field.} was shown to appear in some models of Type IIA string Calabi-Yau compactifications \cite{palti}, but it is not fully clear how general it is and if it applies to any axionic field in string theory. Furthermore, it was shown in \cite{palti} that, at least for the cases studied in the paper, even if the strength of the logarithm is flux-dependent, the proper field distance available before the point at which $\delta s(\phi)>s_0$ and the logarithmic behaviour appears is flux independent and bounded by the Planck mass. Therefore one cannot delay the redefinition of the Kahler metric (and the consequent reduction of the field range) indefinitely by tuning the fluxes. Recently, in \cite{palti2} it was argued that this behaviour is model independent and can be related to the Swampland Conjectures \cite{Vafa:2005ui,Ooguri:2006in}, for which the logarithmic behaviour of the proper field distance at large field is associated to a tower of particles that becomes light for large values of the field. Hence it must be seen as a universal property of a consistent theory of quantum gravity. Here we will analyse this effect in more detail and discuss the features that an effective theory needs to satisfy in order to reproduce this behaviour, with the aim of finding out whether it is accidental or a universal characteristic of string theory. Unfortunatly, the backreaction highly depends on the minimization process and therefore has only been studied in some particular models. 
In this paper we will reformulate the problem in terms of Minkowski 4-forms, which can help to analyse the backreaction from a broader and more model-independent perspective.

 Let us also remark that the backreaction does not have to necessarily spoil inflation, but can have interesting effects. For instance, it can simply correct the inflationary potential leading to interesting flattening effects, e.g. \cite{Buchmuller:2015oma,Bielleman:2016grv}. Here we are interested in studying the viability of having a transplanckian field range in string theory, so we will focus on identifying the cases which give rise to a reduction of the field range.

Let us repeat for convenience the general structure appearing in string compactifications in terms of Minkowski 4-forms,
\beq
V=-Z_{ab}(s^i)F_4^a\wedge *F_4^b +2F_4^a\rho_a(\phi^i)
\label{potII}
\eeq 
Notice that the axions and saxions appear in a very different way in the scalar potential. The former appear within shift invariant functions $\rho(\phi)$ coupled linearly to the 4-forms, while the saxions appear on the kinetic metrics of the 4-forms.
Upon integrating out the 4-forms, the scalar potential becomes
\beq
*F_4^a=Z_{ab}^{-1}(s^i)\rho_b\ \rightarrow \ V=Z_{ab}^{-1}(s^i)\rho_a(\phi^i)\rho_b(\phi^j)
\label{potential}
\eeq
We can then minimize the potential with respect to the saxions, obtaining that the vacuum expectation value for the saxions can always be written as a function of the $\rho(\phi)$ functions. If the metrics $Z_{ab}$ can be written in terms of powers of the saxions, then
\beq
s\sim\frac{\sum_i\rho_1^{n^i_1}\rho_2^{n^i_2}...\rho_a^{n^i_a}}{\sum_j\rho_1^{m^j_1}\rho_2^{m^j_2}...\rho_a^{m^j_a}}
\label{saxion}
\eeq
where the different exponents $(n_a^i,m^i_a)$ can be determined upon minimizing the potential. Remarkably the presence of a possible contribution $V_{loc}(s^i)$ does not break this structure in all the known examples, as we will see below. This procedure is simpler than minimizing the complete scalar potential, since we do not need to know the explicit form of the $\rho$ functions in terms of the axions, but only the form of $Z_{ab}(s)$.
Besides, the backreaction problems appear  intuitively. If the vev of a saxion $s$, which appears in the Kahler metric of the inflaton, is proportional to a $\rho$ function including the inflaton $\phi$, the Kahler metric of $\phi$ will decrease with the inflaton vev, leading to a reduction of the field range. For instance, if $\phi$ appears only in $\rho_1$, all we need is that $n_1-m_1>0$.  Notice also that the metrics $Z_{ab}$ are not arbitrary, but in $N=1$ four dimensional effective theories correspond indeed to the real part of the Kahler metrics of the scalar manifold \cite{Gates:1980ay,Gates:1980az,Deo:1985ix,Binetruy:1996xw,Ovrut:1997ur,Binetruy:2000zx,Cerdeno:2003us,Girardi:2007rs,Nishino:2009zz,Bandos:2011fw,Groh:2012tf,Dudas:2014pva,4forms}. Therefore, once we know the Kahler potential, we can determine $Z_{ab}$ and compute the exponents $(n_a,m_a)$ in eq.\eqref{saxion}. We leave the manifestly supersymmetric description of the above Lagrangian in terms of 3-form supermultiplets for future work. Let us also comment that the fact that the axions appears always within the shift-invariant functions $\rho(\phi)$, implies that the corrections arising from backreaction of the saxions also preserve the discrete shift symmetries of the axions. However, by choosing a branch, the shift symmetry is spontaneously broken and $\rho(\phi)$ can take large values that backreact on the kinetic metric in a non-negligible way.

The question now is how far in field distance we can delay the backreaction effects. The maximum proper field distance available before the inflaton dependence starts dominating the vacuum expectation values of the saxions and the Kahler metric starts depending at leading order  on the inflaton itself is given by
\beq
\Delta\phi = \int \sqrt{K_{\phi,\bar \phi}(s(\phi))} d\phi\sim \phi_c \sqrt{K_{\phi,\bar \phi}(s_0)}
\eeq
\begin{figure}[t]
\begin{center}
\includegraphics[width=9cm]{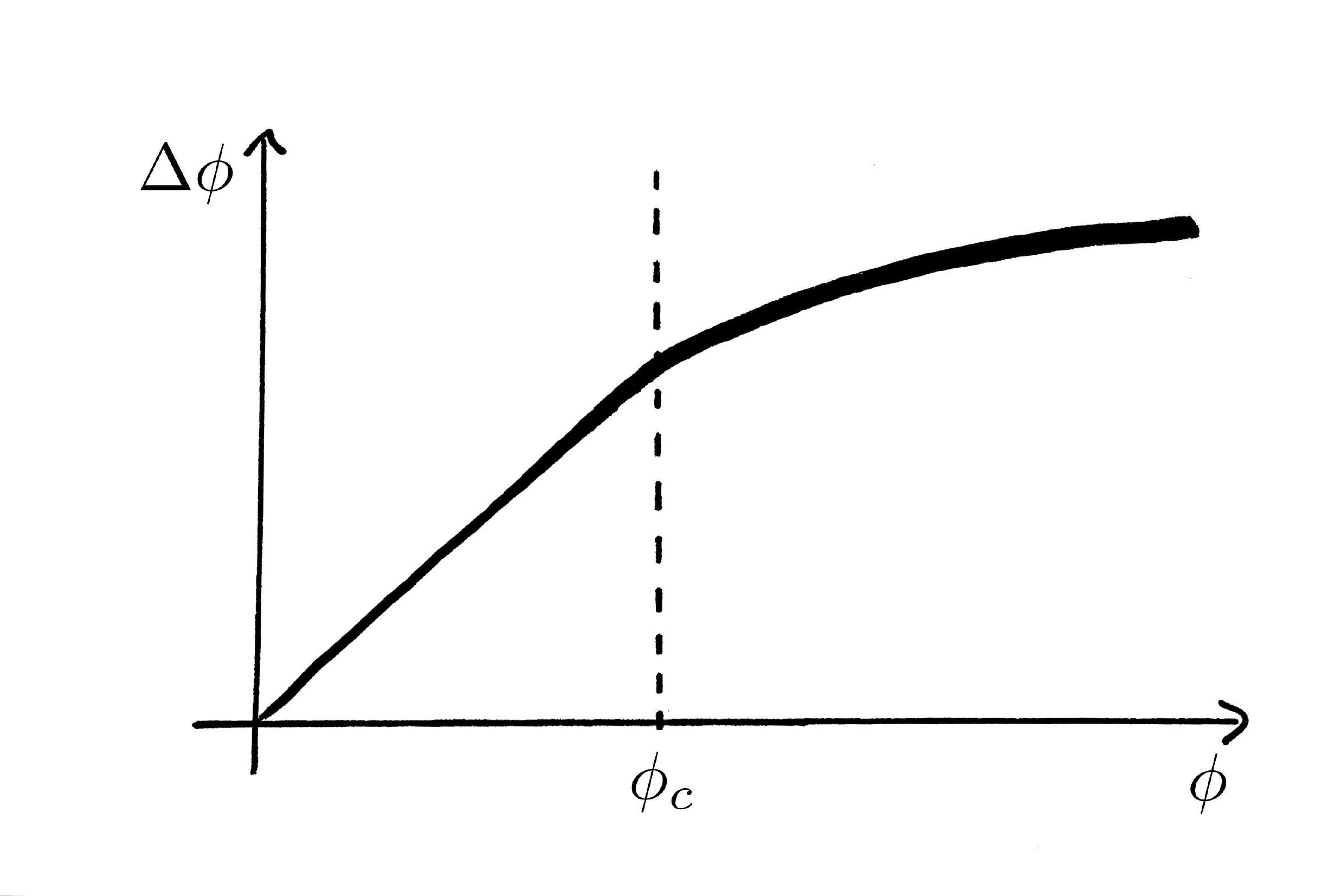}
\end{center}
\vspace{-0.7cm}
\caption{Behaviour of the proper field distance at large field. }
\label{picture}
\end{figure}
where $\phi_c$ is the critical point at which $s(\phi)\sim s_0$ (see figure \ref{picture} for a schematic drawing). Let us recall that this is not the total maximum field range, but only the maximum field range before $s(\phi)> s_0$ and the inflaton dependence on the saxion backreacts on the kinetic metric. Beyond this value, the effective field range will increase at best logarithmically with the field value, as discussed above. Therefore it gives us information about how far we can delay the backreaction effects. The constant value $s_0$ (and consequently $K_{\phi,\bar \phi}(s_0)$) is given by \eqref{saxion} evaluated at the minimum of the potential, i.e. at $\rho=\rho_0$. The critical value $\phi_c$ can be determined by requiring that at least one of the $\rho$ functions including the inflaton dominates over all the constant values $\rho_0$ in \eqref{potential}. However, in order to give quantitative results regarding this issue, we need to know the explicit expression of the shift-invariant $\rho$ functions. Therefore, for concreteness, we will continue the discussion in the next section analysing the closed string sector of Type II string compactifications, where  \eqref{potII} is known \cite{4forms,Carta:2016ynn}.

Before concluding this section, a final remark is in order. The appealing feature of the original Kaloper-Sorbo model is the protection against higher order dimensional operators. However, it has been argued that in more elaborated models (like those coming from string theory) there can be higher operators that break the shift symmetry of the axion, making the formulation in terms of a 4-form useless. We want to stress here that this is not true: there is not explicit breaking of the discrete shift symmetry in any case. All higher dimensional operators correcting \eqref{potII} must preserve the discrete shift symmetry (since it is the remnant of a large gauge transformation) and therefore must appear as functions of the gauge and shift invariant quantities\footnote{Corrections as powers of the field strength $F_4$ give rise to corrections to the scalar potential going as powers of the potential itself, while corrections going as powers of the coupling $F_4\rho$ lead to corrections to the axionic kinetic term involving higher derivative terms of $\phi$ \cite{Kaloper:2016fbr}.} $F_4$ and $\rho$.  The coupling of the axion to the 4-form can be understood as a way to decompactify the compact moduli space of the axion,  converting a circle in configuration space into an helix. 
This process of making global a local symmetry has also been recently discussed in a different context in \cite{Bachas:2016ffl}, and it is the basis underlying axion monodromy. The large gauge transformation relating equivalent vacua requires now to shift the axion and the flux quanta simultaneously. Therefore, upon choosing a vacuum by fixing the flux quanta, the axionic field range can be made bigger than its fundamental periodicity. The backreaction effects can also be understood in this way, without requiring the introduction of any shift symmetry breaking operator that makes the theory out of control. The Lagrangian \eqref{potII} is perfectly shift invariant, and the problems come only upon the choice of vacuum which spontaneously breaks the symmetry. One could think of integrating out the saxions in order to get the effective theory only in terms of the 4-forms and the axions, but this is in general not analytically doable. Besides, it is not clear whether it is even consistent to integrate out the saxion while keeping the 3-form fields in the effective theory. Notice that we are talking about properly integrating out the heavy degrees of freedom and not about making a truncation of the theory by fixing the saxions to their vevs at the minimum of the potential. Therefore we think that the best way of analysing the backreaction effects in axion monodromy is by considering the generalization of Kaloper-Sorbo in \eqref{potII} with the non-trivial kinetic metrics. 
Hence the limitation of Kaloper-Sorbo is not related to mysterious higher dimensional non-shift-symmetric operators, but to the field dependent metrics of the 4-forms and our current ignorance of the manifestly supersymmetric description of Kaloper-Sorbo in $N=1$ supergravity.

\section{Backreaction on the Type IIA(B) closed string sector}

For the sake of concreteness, let us analyse the above backreaction issues in the context of Type II string compactifications, so we can see how the generic properties discussed in the previous section emerge in particular examples.
The four dimensional scalar potential of Type IIA Calabi-Yau orientifold flux compactifications is given by \cite{4forms}
\beq
V=V_{RR}+V_{NS}+V_{loc}
\eeq
where
\begin{multline}
V_{RR}=\frac{e^{K_{cs}}}{2s}\left[-k F_4^0\wedge *F_4^0+2F_4^0\rho_0-4kg_{ij}*F_4^i\wedge F_4^i+2F_4^i\rho_i-\right.\\
\left.-\frac{1}{4k}g_{ij}\tilde F_4^i\wedge *\tilde F_4^j+2\tilde F_4^i\tilde\rho_j+kF_4^m\rho_m\right]
\label{VRR}
\end{multline}
\beq
V_{NS}=e^{K_{cs}}\frac{s^2}{k}H_4\wedge *H_4\ ,\quad V_{loc}=e^{K_{cs}}\frac{2t^3}{3 s t^3}(2 m h_i u^i-2 m h_0 s)
\eeq
and
\beqa
\rho_0=e_0+ b^ie_i-\frac{m}{6}k_{i jk}b^ib^jb^k+k_{i jk}\frac12 q_ib^jb^k-h_0c_3^0- h_ic_3^i\nonumber\\
\rho_i=e_i+k_{ijk}b^jq^k-\frac{m}{2}k_{ijk}b^jb^k\nonumber\\
\tilde\rho_i=q_i-mb_i\nonumber\\
\rho_m=m
\label{rho}
\eeqa
The $\rho$ functions are invariant under the axionic discrete shifts
\begin{align}
&b^i\rightarrow b^i+n^i\ ,\quad q_i\rightarrow \tilde \rho_i(b^i=-n^i)\ ,\quad e_i\rightarrow \rho_i(b^i=-n^i)\ ,\quad e_0\rightarrow \rho_0(b^i=-n^i)\\
&c_3^I\rightarrow c_3^I+n^I\ ,\quad e_0\rightarrow e_0+h_In^I
\end{align}
The 3-form gauge fields come from higher dimensional NSNS and RR p-form fields. In particular,
\beqa
\label{fluxes}
\nonumber F_0=-m\ ,\quad F_2=\sum_iq_i\omega_i\ ,\quad F_4=F_4^0+\sum_ie_i\tilde\omega_i\\ F_6=\sum_iF_4^i\omega_i+e_0dvol_6\ , \quad F_8=\sum_i\tilde F_4^i\tilde \omega_i\ ,\quad F_{10}=F_4^mdvol_6
\label{fluxes}
\eeqa
where $i,a=1,\dots,h_-^{(1,1)}$. The parameters $e_0,e_i,q_i,m$ refer to internal RR fluxes on the Calabi-Yau $Y$, and we get $2h_-^{(1,1)}+2$ Minkowski 4-forms: $F_4^0$, $F_4^i$, $\tilde F_4^i$ and $F_4^m$. The metric is defined as  $g_{ij}=\frac{1}{4k}\int \omega_i\wedge * \omega_j$ .
Similarly the NS $H_3$ background and its $H_7$ dual can be expanded in harmonic forms as
\beq
H_3=\sum_{I=0}^{h_{2,1}^-} h_I\beta_I
\ ,\quad
H_7=\sum_{I=0}^{h_{2,1}^+}  H_4^I\wedge \alpha_I
\eeq
obtaining $h_{2,1}^++1$ additional Minkowski 4-forms $H_4^I$ coming from the NSNS sector.
The 4d axions come from expanding $B_2$ and $C_3$ as follows 
\beq
B_2=\sum_ib_i\omega_i\ ,\quad C_3=\sum_Ic_3^I\alpha_I
\label{B2C3}
\eeq
and correspond to the axionic part of the complex supergravity fields $T,S,U$,
\beqa
T^i=v^i+ib^i\ ,\quad
U^i=u^i+ic_3^i\ ,\quad
S=s+ic_3^0
\label{TSU}
\eeqa
Upon integrating out the 3-form fields via their equations of motion,
\beqa
*_4F_4^0=\frac1k \rho_0\ , \quad
*_4F_4^i=\frac{g^{ij}}{4k}\rho_j\nonumber\\
*_4\tilde F_4^i=4kg^{ij}\tilde \rho_j\ , \quad
*_4F_4^m=\rho_m
\label{F4m}
\eeqa
the RR and NS scalar potential reads
\beq
V_{RR}+V_{NS}=\frac{e^{K_{cs}}}{s}\left[\frac{1}{2k}|\rho_0|^2+\frac{g_{ij}}{8k}\rho^i\rho^j+2kg_{ij}\tilde\rho^i\tilde\rho^j+k|\rho_m|^2+\frac{1}{k}c_{IJ}\rho^{I}\rho^{J}\right]
\eeq
where $c_{IJ}$ is the metric in the complex structure moduli space.
By plugging eq.\eqref{rho} into the above formula one recovers the well known scalar potential of Type IIA flux compactifications \cite{Grimm:2004ua,Louis:2002ny,Villadoro:2005cu,DeWolfe:2005uu,Camara:2005dc}. 
We will use this scalar potential, written as a sum of the squares of the different $\rho$ functions, to analyse the backreaction of the saxions. 

Finally, the kinetic metrics are determined by second derivatives of the Kahler potential, which is given by \cite{Grimm:2004ua}
\beq
K=-2log(\mathcal{F}_{KL}(N^K-\bar N^K)(N^L-\bar N^L))-log(k_{ijk}(T+T^*)^i(T+T^*)^j(T+T^*)^k)
\eeq
Here $N^K$ stands both for the dilaton $S$ and complex structure moduli $U$ with $K=0,\dots ,h^{2,1}$, while $\mathcal{F}_{KL}=\partial_K\partial_L \mathcal{F}$ is the second derivative of the prepotential $\mathcal{F}$ inherited from the $N=2$ unorientifolded theory. Once we are located at a special point in the complex structure moduli space we can expand the periods around it and obtain an explicit expression for the Kahler potential. For instance, if we consider a manifold with only one complex structure modulus, the result near the large complex structure point will be given by 
\beq
K=-log(S+S^*)-3log(U+U^*)-log(k_{ijk}(T+T^*)^i(T+T^*)^j(T+T^*)^k)
\eeq
For concreteness, we will use this Kahler potential in the following, and leave the study of other special points in the complex structure moduli space for future work.

\subsection{Minima of the potential}

First we minimize the potential with respect to the axions. Since they only appear inside the $\rho$ functions we do not need to worry about $V_{loc}$. If all the fluxes are not vanishing, the axions $b^i$ and the combination $\phi=h_0c_3^0+ h_ic_3^i$ are stabilized by the fluxes. Since $\phi$ appears only inside $\rho_0$ which appears quadratically in the potential, minimazation with respect to $\phi$ implies $\rho_0=0$ at the minimum,
\beq
\rho_0=0 \rightarrow \phi_0=e_0+\frac{q}{m}(\gamma-\frac{q^2}{6m})
\label{rho0}
\eeq
where we denote $\gamma=e_i+\frac12 \frac{q^2}{m}$. Furthermore, each $\rho$ function is the derivative with respect to $b^i$ of the previous one, obtaining
\beq
\frac{\partial V}{\partial b^i}=\frac{2}{su^3}\left[\frac{1}{2k}\rho_0\rho_i+\frac{g^{jk}}{8k}k_{ijl}\rho_k\tilde \rho^l+2kg_{ij}\tilde\rho^j\rho^m\right] 
\eeq
We have therefore two options,
\beqa
(I)\quad \tilde\rho^i=0 \rightarrow b^i_0=q^i/m\\
(II) \quad k_{ijl}\frac{g^{jk}}{8k}\rho_k+2kg_{il}\rho^m=0
\eeqa
For simplicity let us consider only one Kahler modulus $t$ (ie, $h^{(1,1)}_-=1$) and one complex structure modulus $u$, which will be enough for our purposes. Notice that then the index $i$ is not a running index but just $i=1$. We will leave it, however, as a label to distinguish the different $\rho$ functions.  We also denote the NS fluxes  as $\rho_{h_0}=h_0$ and $\rho_{h_i}=h_i$. The scalar potential is then given by 
\beq
V_{RR}+V_{NS}=\frac{4}{st^3u^3}\left[(\rho_0)^2+\frac{t^2}{3}(\rho_i)^2+\frac{3t^4}{36}(\tilde\rho_i)^2+\frac{t^6}{36}(\rho_m)^2+s^2\rho_{h_0}^2+u_i^2\rho_{h_i}^2\right]
\eeq
and the two possible solutions become
\beqa
(I)\quad &\tilde \rho_i=0 \rightarrow b=q/m\\
&\rho_i=\gamma
\\
(II) \quad & \rho_i=-\frac{t^2\rho_m}{4}\rightarrow b_0=\-q/m\pm m^{-1/2}\sqrt{2\gamma-t^2/2}\\
&\tilde \rho_i=\pm m^{1/2}\sqrt{2\gamma-t^2/2}
\label{sets}
\eeqa
Next we minimize the potential with respect to the saxions. It can be easily checked that both sets of solutions (since both will satisfy $\rho_i=\gamma$) yield (up to $\mathcal{O}(1)$ factors)
\beqa
\quad s_0\sim \frac{\rho_i^{3/2}}{\rho_{h_0}\sqrt{\rho_m}}\ ,\quad
u_0\sim \frac{\rho_i^{3/2}}{\rho_{h_1}\sqrt{\rho_m}}\ ,\quad
t_0\sim \frac{\rho_i^{1/2}}{\sqrt{\rho_m}}
\eeqa
The only difference is that in the first set of solutions $\tilde \rho_i=0$ (implying $b=q/m$) while in the second set $\tilde\rho_i=\sqrt{6\gamma m}$ (implying  $b=\frac{q}{m}\pm \sqrt{\frac{6\gamma}{m}}$). 
Notice that the result can be written indeed as a ratio of the different $\rho$ functions, in agreement with section 3. The explicit parametric dependence on the fluxes is then given by
\beqa
s_0\sim \frac{\gamma^{3/2}}{h_0\sqrt{m}}\ ,\quad
u_0\sim \frac{\gamma^{3/2}}{h_1\sqrt{m}}\ ,\quad
t_0\sim \frac{\gamma^{1/2}}{\sqrt{m}}
\label{t0}
\eeqa
It is also interesting to notice how the structure of minima found for instance in \cite{Camara:2005dc} appears in a more transparent and clear way in terms of the $\rho$ functions.

\subsection{Backreaction on the kinetic metric}

Let us study now  how these minima change when we displace the inflaton away from its minimum. We will consider two options, depending on whether the inflaton belongs to the NS or the RR sector.\\

\noindent\textbf{RR sector}

The inflaton is the linear combination of RR axions $\phi=h_0c_3^0+ h_ic_3^i$. When this field is displaced away from its minimum \eqref{rho0}, then $\rho_0\neq 0$, and the minima will depend both on $\rho_i=\gamma$ and $\rho_0$. In fact, for large field\footnote{Actually, for large field the saxion vevs will go as $\rho_0(\phi)/h$, so the inflaton appears always within the shift invariant $\rho$ functions. This is consistent with the fact that any correction to the effective theory cannot break explicitely the shift symmetry.}
\beqa
s\underset{\phi\to \infty}{\longrightarrow} \frac{\phi}{h_0}\ ,\quad
u\underset{\phi\to \infty}{\longrightarrow}  \frac{\phi}{h_1}
\eeqa
and the physical field distance scales as
 \beq
\Delta\phi=\int K_{\phi,\bar \phi}^{1/2} d\phi\sim \int \sqrt{\frac{1}{h_0^2s(\phi)^2}+\frac{1}{h_1^2u(\phi)^2}}d\phi\underset{\phi\to \infty}{\longrightarrow} \int \frac{1}{\phi}d\phi\sim \ log(\phi)
\eeq
ruling out parametrically large field ranges. The larger the field displacement is, the smaller is the physical increase in field distance. Notice also that how fast the Kahler metric decreases at large field is model dependent (it might depend on the flux parameters). Therefore, large field inflation with a field displacement of a few times the Planck mass is still possible by tuning the fluxes. However, here we are not interested in whether a modest large field excursion is allowed, but in understanding if string theory distinguishes between sub- or trans-planckian field distances. We have seen that parametrically large field values are not possible, but does anything special happen when crossing the threshold $\Delta\phi\sim M_p$? In order to answer this question, let us see how far we can delay the appearance of the logarithm in the above expression.

Let us roughly divide the field range in two regimes, before and after the metric becomes field-dependent on the inflaton,
\beq
\Delta\phi \sim \int_{\phi_0}^{\phi_c} K_{\phi,\bar \phi}^{1/2} d\phi+ \int_{\phi_c}^{\phi} K_{\phi,\bar \phi}^{1/2} d\phi\sim\Delta\phi_-+ \Delta\phi_+
\eeq
where $\phi_c$ is the critical value at which $\delta s(\phi_c)\sim s_0$. Before this point, the metric is roughly constant (ie, it does not depend to leading order on the value of $\phi$) and the field range is given by
\beq
\Delta\phi_- \sim (\phi_c-\phi_0)K_{\phi,\bar \phi}^{1/2} \sim (\phi_c-\phi_0)\sqrt{\frac{1}{h_0^2s_0^2}+\frac{1}{h_1^2u_0^2}}
\eeq
After $\phi_c$ the leading contribution on the metric will depend on $\phi$ itself and $\Delta\phi_+$ will grow at best logarithmically with the field value. The question is, how large can $\Delta\phi_-$ be? As pointed out in \cite{palti} this quantity is flux-independent in the above Type IIA compactifications and cannot be bigger than the Planck mass, as we proceed to explain in the following. Let us consider for simplicity $h_0>h_1$ without loss of generality, so we can focus only on the backreaction of the dilaton field.
 For small displacements, the dilaton minimum is modified as follows,
\beq
s=s_0+\mathcal{O}(1)\frac{\sqrt{m}}{h_0\gamma^{3/2}}\rho_0^2=s_0(1+\frac{\phi_0^2}{h_0^2s_0^2})
\eeq
so $\delta s(\phi)>s_0$ when $\phi>\phi_c\sim h_0s_0$. Therefore,
\beq
\Delta\phi_-  \sim \frac{\phi_c-\phi_0}{h_0s_0}\lesssim 1
\eeq
in Planck units. This result can also be inferred by looking directly at the scalar potential. At the minima of the potential, all the vacuum expectation values for the saxions depend on the quantity $\rho_i=\gamma$. When the inflaton is displaced away from its minimum, $\rho_0\neq 0$ and the minimization process depends on the balance between the terms involving $\rho_0$ and $\rho_i$. The inflaton dependence will  dominate when $\rho_0>t\rho_i$, implying $\phi>\phi_c\sim \gamma^{3/2}/\sqrt{m}\sim h_0 s_0$, recovering the previous result.

To sum up, if $\phi_c$ and $s_0$ scale the same way on the flux parameters, $\Delta\phi_-$ is flux independent and at best of order  $\Delta\phi_-\sim \phi_c/s_0\sim 1$ in Planck units. This implies that when the field displacement becomes transplanckian, the metric starts depending on the field value itself in a non-negligible way, such that it decreases when the field increases. The flux-independence of the available field range before backreaction effects become important  suggests that there is a qualitative change on the effective theory when crossing $\Delta\phi\sim M_p$.\\

\noindent\textbf{NS sector}

Let us analyse now if the same behaviour appears when the inflaton is a NS axion $b_i$. In this case, the physical field distance will be given by
\beq
\Delta b \sim \int K_{T\bar T}^{1/2} db \sim\int \frac{\sqrt{3}}{t(b)}db
\eeq
When $b$ travels away from its minimum, the variation on the $\rho$ functions is given by
\beq
\Delta \rho_a=-m\delta b\ ,\quad
\Delta \rho_i=-\frac{m}{2}(\delta b)^2\ ,\quad
\Delta \rho_0=f \delta b-\frac{m}{6}(\delta b)^3
\eeq
We should replace this into the scalar potential and minimize again to see how the new minimum for the kahler modulus depends on $b$. However, the system of equation becomes non-linear and cannot be resolved analytically. Numerical studies show that the kinetic metric again decreases with the field value, and that $\Delta b_-$ seems to be roughly independent of the fluxes. Let us though remark that in order to compute $\Delta b_-$ we do not need to minimize the scalar potential for arbitrary values of $b$. We can estimate the critical value $b_c$ by simply requiring that $\Delta\rho_i > \rho_i(b_0)$, i.e. that the variation on $\rho_i$ dominates over the value of $\rho_i$ at the minimum\footnote{The same result can be obtained by requiring $t^2\Delta\rho_a>\rho_i$ as in the previous section, so the term involving $b$ dominates over the constant terms in the scalar potential.}. This clearly remarks the simplicity of the computations when writing the potential in terms of 4-forms. This implies that
\beq
b_c\sim \sqrt{\frac{\gamma}{m}}
\eeq
which is valid for both sets of solutions in \eqref{sets}, since for both of them $\rho_i(b_0)=\gamma$. Notice that $b_c$ has the same parametric dependence on the fluxes than the constant vev $t_0$ \eqref{t0}, implying that the maximum physical field distance available before the metric starts decreasing with the field vev,
\beq
\Delta b_-=\frac{b_c}{t_0}\sim 1
\eeq
is again flux-independent and at best of the order of the Planck mass, in agreement with \cite{palti}.\\

\subsection{Remarks and comments}
The flux independence of $\Delta\phi_-$ is clearly remarkable. If it is not accidental but can be proved to be a generic property of any string model, it would be a strong hint in favour of a fundamental reason why transplanckian field ranges are disfavoured in string theory (see \cite{palti,palti2} for the relation to the Swampland Conjecture). 
Let us therefore take a step back and analyse the reason of this miraculous cancellation. Consider that the kinetic metric of the inflaton is inversely proportional to a saxion $s$ whose vacuum expectation value is given by
\beq
s=s_0+\delta s(\phi)
\eeq
with $\delta s(\phi)$ vanishing at the minimum of $\phi$. We have seen that $\delta s(\phi)$ will be given by some ratio of the $\rho$ functions. In the case in which this ratio is proportional to $\phi$, the kinetic metric will decrease when increasing $\phi$ leading to a reduction of the effective field range. In terms of the parameters of the effective theory, this ratio is proportional to some function of the mass ratio
\beq
\delta s(\phi)\propto f\left(\frac{m_\phi}{m_s}\phi\right)
\eeq
so that a big mass hierarchy suppresses the backreaction. In the limit $m_s\to \infty$ the saxion is frozen at its minimum value and the kinetic metric for $\phi$ is not field dependent anymore (we would need infinite energy to move $s$ from its minimum). However, our results and those from \cite{palti} show that a big mass hierarchy is not enough to suppress the backreaction effects and send the logarithmic behaviour far away in field distance. One needs to check that by decreasing $\frac{m_\phi}{m_s}$ one is not increasing in the same way the value of $s_0$, so both effects cancel each other. In the examples studied above, one cannot vary the mass hierarchy while keeping the vev $s_0$ constant. In fact, both scale in the same way with the fluxes (due to the fact that $\phi_c \sim s_0$) implying that the maximum proper field distance available before reaching the inflection point ($\delta s(\phi) \sim s_0$) is flux-independent and of order one in Planck units. Can we then construct a string model in which $s_0$ and $\phi_c$ vary independently? We have shown that this is not possible within the closed string sector of Calabi-Yau compactifications of Type IIA. However, the open string sector might provide a counter-example to this behaviour, as we will explain in the next section.

Before moving to the open string sector, let us roughly discuss the case of Type IIB. The effective theory of IIB Calabi-Yau orientifold compactification with fluxes in terms of Minkowski 4-forms was derived in \cite{4forms}. However, the Minkowski 4-forms there correspond to the complexified 4-forms $G_4$ coming from expanding $G_7$ in harmonic internal 3-forms. Here $G_7$ is the dual in ten dimensions of $G_3$,
\beq
G_7=*_{10}G_3\ ,\quad G_3=F_3-S H_3=\tilde F_3-ie^{-\phi} H_3
\eeq
with $\tilde F_3=F_3-C_0 H_3$. Therefore $G_7$ already includes the dilaton in its definition.
 In order to get the structure of \eqref{potII} we need to reformulate the effective theory in terms of the real 4-forms $\tilde F_4$ and $H_4$. In this case, since neither axions coming from $B_2,C_2$ nor $C_4$ (belonging to the Kahler sector) get stabilized by fluxes 
 \cite{Ibanez:2012zz}, the only candidate for axion monodromy (without considering extra contributions from the open string sector) is the fundamental axion $C_0$, which will appear within the $\rho$ function coupled to the RR gauge invariant form $\tilde F_4$.  The structure is then much simpler than in IIA, since everything will depend on this particular $\rho$ function making the backreaction effects unavoidable. Another option would be to consider the inflaton within the complex structure moduli sector. Interestingly, axions with exact discrete shift symmetries (and approximate continuous symmetries) appear near special points in the complex structure moduli space that admit discrete monodromy symmetries of infinite order \cite{Garcia-Etxebarria:2014wla}. It would be interested to study if an analogous formulation in terms of effective 4-forms is possible in these cases as well. In any case, we do not expect qualitatively different results from Type IIA, at least near the large complex structure point, because many models will be  related by those of IIA by mirror symmetry. However, the complex structure moduli space is richer than considering only the large complex structure point, and different effective theories can arise near different special points \cite{Garcia-Etxebarria:2014wla,Blumenhagen:2016bfp,Bizet:2016paj}. Therefore, even if it is difficult to imagine how one can get a qualitatively different behaviour for the field range at large field, we cannot discard that possibility and leave a more careful analysis for future work. Instead, here we will focus now on the open string sector, which looks more promising to overcome the difficulties discussed above.

\section{A possible way out: Open string moduli}

The introduction of the open string sector will have two effects in the effective theory. First, new superpotential couplings can be generated, which in terms of Minkowski 4-forms is equivalent to modify the $\rho$ functions in order to include the open string axions inside. Besides, if one turns on open string fluxes, new dual Minkowski 4-forms will appear in the effective theory. Here we will not  write the full scalar potential including the new open string fields. For our purposes it is enough to consider that the structure \eqref{VII} is still preserved. A first analysis of the potential in terms of 4-forms in the case of Type IIA with D6-branes was performed in \cite{Carta:2016ynn} (see also \cite{Ibanez:2014swa,4forms,Grimm:2008dq} for partial results for D7-branes and D5-branes). However, the complete reformulation of the scalar potential in terms of both closed and open 4-forms is still missing. The second effect is the modification of the Kahler potential due to the redefinition of the holomorphic chiral variables \cite{Grimm:2011dx,Kerstan:2011dy,Camara:2011jg,Marchesano:2014bia,Grimm:2010ks,Jockers:2004yj,Jockers:2005zy,Corvilain:2016kwe}. In Type IIA compactifications, the complex structure moduli is redefined in the presence of D6-brane moduli \cite{Grimm:2011dx,Kerstan:2011dy,Camara:2011jg,Marchesano:2014bia,Grimm:2010ks} and the Type IIA Kahler potential for the complex structure sector becomes
\beq
K_{IIA}=-2log(\mathcal{F}_{KL}(N^K-\bar N^K-Q^K(\phi-\bar\phi)^2)(N^L-\bar N^L-Q^K(\phi-\bar\phi)^2))
\eeq
where $\phi$ are D6 Wilson lines. Analogously, in Type IIB the D7 position moduli appear modifying the dilaton \cite{Jockers:2004yj,Jockers:2005zy}. Due to the lower codimension of the D7-brane, the backreaction of the D7 cannot be treated in a consistent global way within perturbative Type IIB string theory. Therefore, it is better to continue the analysis in the framework of F-theory. The complex structure deformations of the Calabi-Yau 4-fold can then be identified in the weak coupling limit with the IIB axio-dilaton, bulk complex structure moduli and positions of the D7's. The Kahler potential of the low energy theory is given by \cite{Jockers:2004yj,Grimm:2010ks}
\beq
K_{CS}=-log(\int_X \Omega_4\wedge \bar \Omega_4)\overset{g_s\to 0}{\approx} -log((S_0-\bar S_0)\Pi_i(z)Q^{ij}\bar \Pi_j(\bar z)+f(z,\bar z,\phi,\bar\phi)))
\eeq
where $\phi$ are the D7 position moduli and $\Pi_i(z)$ the period vectors of the base depending on the bulk complex structure coordinates $z$. In particular, the Kahler potential for the vector moduli of F-theory compactified on $K3\times\tilde K3$ at the orientifold point (strong coupling limit of Type IIB on $K3\times T^2/Z_2$) is known \cite{GarciaEtxebarria:2012zm} to all orders in $g_s$ and perturbatively exact in $\alpha'$. At zeroth order in $\alpha'$ (but exact in $g_s$) the Kahler potential for the vector moduli is simply given by
\beq
K=-log((S+\bar S)(U+\bar U)-\frac12 \sum_a(\phi^a+ \bar\phi^a)^2)-log(T+T^*)
\label{K3}
\eeq
where $S$ and $T$ are the complex structure and Kahler modulus of the $T^2$ on the base respectively. 
Higher orders in $\alpha'$ will imply a mixing between the Kahler and complex structure sectors. This Kahler potential also corresponds to the one appearing in toroidal compactifications of Type IIB, with $U$ being the complex structure coordinate along the torus transverse to the brane.  

We have focused on D6 wilson lines and D7 position moduli because they can be stabilized by fluxes and are therefore good candidates for monodromic axions. Notice that the leading order contribution (tree level in $\alpha'$ and $g_s$) for the Kahler metric of open string fields depends only on the closed string saxions. This opens a new possibility: if one can separate the source of stabilizing closed and open string moduli, one could a priori increase the critical value $\phi_c$ while keeping the saxion vev $s_0$ constant. 

In the case of D6-branes, this can be done by turning on a magnetic worldvolume flux $n_F$ on the brane. This flux will affect to the stabilization of both closed and open string fields, while some of the RR closed string fluxes can be chosen in order not to affect the open string field identified with the inflaton. Therefore we can tune $n_F$ to small values to increase $\phi_c$ while keeping $s_0$ approximately constant (since the leading order contribution will still come from closed string fluxes). However, the possible choices of fluxes that are globally consistent and satisfy the 10d equations of motion of the internal manifold are not well understood, because the internal geometry is not Calabi-Yau anymore when we add the fluxes. Therefore it is not clear for us if the tuning of fluxes required to suppress the backreaction effects is even globally allowed. In this sense, it is better to consider the analogous model in Type IIB with D7's, where the global constraints are better understood. 

In the case of D7's, one can add a supersymmetric ISD (2,1) flux  to stabilize the D7 position moduli, which will appear as an extra term in the superpotential. When this flux is set to zero, a real field $\phi^-$ parametrize one of the transverse directions to the brane becomes massless, while the closed string saxions and the scalar $\phi^+$ parametrize the other direction can be kept stabilized by the presence of other fluxes like ISD (0,3)-fluxes (which break supersymmetry). The position modulus which is stabilized even if the there is not explicit term in the superpotential involving the open string fields is the one appearing in the Kahler potential, due to the well known Giudice-Masiero mechanism. In the notation of \eqref{K3}, it corresponds to $\phi^+=\phi+\phi^*$, while $\phi^-=\phi-\phi^*$ would be the inflaton. Therefore we can play the same game and tune the (2,1)-flux to increase $\phi^-_c$ while keeping $s_0$ (and $\phi^+_0$) approximately constant. Since ISD fluxes satisfy the supergravity ten-dimensional equations of motion of a Calabi-Yau manifold \cite{Giddings:2001yu}, this choice of fluxes is also consistent globally. Notice also that we are assuming that the transverse space to the D7 admits periodic directions, along the lines of \cite{Arends:2014qca,Hebecker:2014eua,Ibanez:2014swa}. This example will be analysed in detail in \cite{new} which presents the implementation of Higgs-otic inflation \cite{Ibanez:2014kia,Ibanez:2014swa,Bielleman:2015lka,Bielleman:2016grv} in a moduli stabilization framework. Therefore we refrain ourselves from giving more technical details here. We will instead discuss the concrete properties and assumptions which allow this behaviour. 

Notice that the above examples rely on the fact that the kinetic metric of the inflaton to leading order does not depend on the scalar within the same supermultiplet than the inflaton, but on a scalar from a different sector (let us denote it $s_0$) which can be stabilized independently and whose vev remains approximately constant even if the inflaton mass tends to zero. This is equivalent to engineer a sort of hierarchy between the inflationary potential and the scalar potential of the saxion $s_0$, which is possible if there is a flux entering in one of them but not in the other one. The fact that this seems to be possible only upon introducing an open string sector nicely fits with the results of \cite{Blumenhagen:2014nba,Hebecker:2014kva}, in which this kind of hierarchy was not possible within the complex structure moduli space of a Calabi-Yau threefold but might be realisable in Calabi-Yau fourfolds (from the point of view of M/F-theory the D7 position moduli are included as part of the complex structure deformations of the $CY_4$). In the example of D7-branes discussed above, we have:
\begin{itemize}
\item (2,1)-fluxes: They contribute to stabilize both closed string moduli (including $s_0$) and D7 position moduli (including the inflaton $\phi^-$). This implies that the saxion vev will depend on the inflaton field and eventually  backreact on the kinetic metric of the inflaton, leading to the logarithmic behaviour $\Delta\phi \sim log(\phi)$ for large field.
\item (0,3)-fluxes: They contribute to stabilize the closed string moduli and one D7 position modulus, but do not affect to the stabilization of the inflaton field $\phi^-$ (which in the absence of (2,1)-fluxes is a flat direction). This implies that $\phi_c$ and $s_0$ do not have the same dependence on the fluxes. For instance, if the contribution from (0,3)-fluxes is more important than the one from (2,1)-fluxes, then one can increase $\phi_c$ while keeping $s_0$ approximately constant (at the value determined by the (0,3) fluxes). Therefore the field range available before the logarithmic behaviour becomes important is flux-dependent and one can a priori delay the backreaction effects beyond the Planck mass in proper field distance.
\end{itemize}

Notice that the same arguments work for the backreaction coming from the open string fields in the Kahler potential \eqref{K3}, since they are also stabilized by (0,3)-fluxes. Besides, if $\phi^+$ becomes larger than $\sqrt{s_0u_0}$, the function inside the logarithm in \eqref{K3} becomes negative and we exit the Kahler cone. This implies that our set of coordinates is not valid anymore in the local patch under consideration and we have to redefine the complex structure moduli to reabsorb the shift on the open string modulus. Therefore, what physically matters is the backreaction from the combination $su-(\phi^+)^2$. 


Let us remark that the above example is possible because the source of supersymmetry breaking gives the leading contribution to stabilize the saxions while has no effect on the inflaton field mass. Since we are breaking supersymmetry, one should worry if loop or higher order corrections do not spoil this behaviour. However, we have argued that the axion appears in the Lagrangian only inside shift-invariant functions $\rho(\phi)$, and higher order corrections must appear as functions of $\rho(\phi)$. This leads to a sort of protection (similar to a chiral symmetry for fermionic masses) in the sense that the corrections remain naturally small if $\rho(\phi)$ was initially small. If one was able to find a source of supersymmetry breaking which stabilizes the axion (the inflaton) but not the saxion, then one could hope to get rid also of the backreaction effects which induce the logarithmic behaviour at large field. However, the saxions do not enjoy such a protection coming from the 3-form fields, so higher order corrections will completely spoil any hierarchy generated at leading order. This is the reason why we expect that the backreaction effects yielding that the proper field distance grows at best logarithmically at large field is a generic feature of string theory, but one can still delay this backreaction effects by tuning the fluxes in some cases. This would imply that parametrically large field values are forbidden in a consistent theory of quantum gravity, but the constraint on the field range is not necessarily tied to the Planck mass in axion monodromy models.


These properties suggest that models of axion monodromy based on open string fields might provide counterexamples to the universal behaviour of $\Delta\phi_-$ discussed in the previous section (and previously in \cite{palti}), since here $\Delta\phi_-$ is flux-dependent. Therefore the threshold at which backreaction effects become important can be set to a transplanckian value. 
However, a more careful analysis is required before extracting general conclusions, to check that we are not missing any relevant backreaction effect that could also reduce the field range and forbid transplanckian excursions. 


\section{Conclusions}

We have analysed the differences between the original Kaloper-Sorbo description of axion monodromy and the effective theory for axions arising in $N=1$ four-dimensional string compactifications of Type IIA/B. The latter can be completely reformulated in terms of a supersymmetric generalization of Kaloper-Sobro with: non-linear couplings to the 4-forms, multiple 4-forms and multiple axions, and field-dependent kinetic metrics of the 3-form fields (depending on the non-periodic scalars of the compactification). The Minkowski 4-forms couple to shift invariant functions $\rho(\phi)$ which encode all the dependence on the axions in the effective theory. Since the axions do not necessarily appear linearly in $\rho(\phi)$ this gives rise, not only to mass terms, but to more general couplings in the effective scalar potential. The discrete shift symmetry of the axions can only be broken spontaneously, so there cannot be higher dimensional operators generating a explicit breaking. This is translated into the fact that all higher order corrections must appear as functions of the gauge-invariant 4-form field strength. However, the presence of multiple 4-forms imply that the higher order corrections can appear as functions of combinations of the different 4-forms mixing different parts of the potential, and not simply as powers of the leading order potential itself. Finally, the presence of the non-periodic scalars (saxions) in the kinetic metric of the 3-form fields leads to the backreaction issues that make axion monodromy models technically involved and difficult to control. By displacing the inflaton away from its minimum one can also destabilize the saxions from their corresponding minima, which might lead to non-negligible modifications of the effective theory and the inflationary dynamics.

We have analysed these backreaction issues in terms of the Minkowski 3-form fields. In particular, we have focused on the case in which the saxions backreact on the Kahler metric of the inflaton, inducing a redefinition of the canonical field and therefore of the proper field distance. If the vev of a saxion (appearing in the Kahler metric of the inflaton) is, upon minimization of the potential, proportional to the inflaton vev, then the proper field distance (for the inflaton) will grow at best logarithmically with the inflaton for large field values. Whether the saxion depends on the inflaton vev is easy to check when writing the effective theory in terms of Minkowski 4-forms, where axions and saxions enter in the potential in a very different way. In particular, the vev of the saxions will be given by ratios of the different $\rho(\phi)$ functions. Therefore, in order to see if the saxion depends on the inflaton we do not need to minimize the full scalar potential but we only need to have information about the metrics of the 3-form fields and about in which $\rho(\phi)$ appears the inflaton (i.e. to which 4-form the inflaton couples). Furthermore, the metric of the 3-form fields is not arbitrary but determined by the Kahler potential of the scalar manifold in $N=1$ compactifications. The question is therefore whether string theory allows for Kahler metrics free of these backreaction problems.

We have then studied in detail  the case of flux compactifications of Type IIA/B in orientifold Calabi-Yau manifolds. We find, in agreement with \cite{palti}, that any axion within the closed string sector will suffer from these backreaction effects and its proper field distance will scale at best logarithmically at large field. This nicely fits with the reluctance of string theory to get parametrically large field displacements. But even more intriguing, the maximum field range before these backreaction effects become important turns out to be flux-independent and tied to the Planck mass. If this behaviour was universal for any string model it would point to a fundamental obstruction to have a transplanckian field range in axion monodromy models within string theory.

We propose, however, some possible counterexamples to this universal behaviour based on open string fields. In particular, one can consider Type IIB/F-theory compactifications with D7-branes, where one can have two sets of fluxes which allow for a sort of mass hierarchy between the saxions and the inflaton (the latter belonging to the open string sector). In these models, one can tune the fluxes in such a way that even if the logarithmic behaviour at large field is unavoidable, it can be delayed away in field distance. Therefore the backreaction effects are not necessarily tied to the Planck mass. 
 It would not be the first time that the introduction of the open string sector leads to new features that are not available within the closed string sector of perturbative Type II string theory. For instance, the prototypical mechanisms to get deSitter vacua in string theory require the presence of anti-D3 branes or D7-branes with fluxes. 
 But as always happen when we add new ingredients, the model becomes more involved and a more careful analysis is required to check the global consistency of the proposal and ensure that we are not missing any relevant backreaction effect. In any case, they look like a promising arena to test if string theory only disfavours parametrically large displacements or instead a fundamental obstruction appears as soon as the field range becomes transplanckian. In addition to the implications for large field inflation, the definition of the boundaries of the string landscape is interesting by itself and clearly deserves more investigation.

\section*{Acknowledgments}

We thank Luis Ib\'a\~nez, Fernando Marchesano, Miguel Montero, Eran Palti, Diego Regalado and Clemens Wieck for useful discussions. This work is supported by  a grant from the Max Planck Society. 


\bibliography{refs}

\providecommand{\href}[2]{#2}\begingroup\raggedright\begin{thebibliography}{100}

\bibitem{Silverstein:2008sg}
E.~Silverstein and A.~Westphal, ``{Monodromy in the CMB: Gravity Waves and
  String Inflation},'' {\em Phys. Rev.} {\bf D78} (2008) 106003,
\href{http://www.arXiv.org/abs/0803.3085}{{\tt 0803.3085}}.

\bibitem{McAllister:2008hb}
L.~McAllister, E.~Silverstein, and A.~Westphal, ``{Gravity Waves and Linear
  Inflation from Axion Monodromy},'' {\em Phys. Rev.} {\bf D82} (2010) 046003,
\href{http://www.arXiv.org/abs/0808.0706}{{\tt 0808.0706}}.

\bibitem{Flauger:2009ab}
R.~Flauger, L.~McAllister, E.~Pajer, A.~Westphal, and G.~Xu, ``{Oscillations in
  the CMB from Axion Monodromy Inflation},'' {\em JCAP} {\bf 1006} (2010) 009,
\href{http://www.arXiv.org/abs/0907.2916}{{\tt 0907.2916}}.

\bibitem{Berg:2009tg}
M.~Berg, E.~Pajer, and S.~Sjors, ``{Dante's Inferno},'' {\em Phys. Rev.} {\bf
  D81} (2010) 103535,
\href{http://www.arXiv.org/abs/0912.1341}{{\tt 0912.1341}}.

\bibitem{Pajer:2013fsa}
E.~Pajer and M.~Peloso, ``{A review of Axion Inflation in the era of Planck},''
  {\em Class. Quant. Grav.} {\bf 30} (2013) 214002,
\href{http://www.arXiv.org/abs/1305.3557}{{\tt 1305.3557}}.

\bibitem{Gur-Ari:2013sba}
G.~Gur-Ari, ``{Brane Inflation and Moduli Stabilization on Twisted Tori},''
  {\em JHEP} {\bf 01} (2014) 179,
\href{http://www.arXiv.org/abs/1310.6787}{{\tt 1310.6787}}.

\bibitem{Palti:2014kza}
E.~Palti and T.~Weigand, ``{Towards large r from [p, q]-inflation},'' {\em
  JHEP} {\bf 04} (2014) 155,
\href{http://www.arXiv.org/abs/1403.7507}{{\tt 1403.7507}}.

\bibitem{Marchesano:2014mla}
F.~Marchesano, G.~Shiu, and A.~M. Uranga, ``{F-term Axion Monodromy
  Inflation},'' {\em JHEP} {\bf 09} (2014) 184,
\href{http://www.arXiv.org/abs/1404.3040}{{\tt 1404.3040}}.

\bibitem{Blumenhagen:2014gta}
R.~Blumenhagen and E.~Plauschinn, ``{Towards Universal Axion Inflation and
  Reheating in String Theory},'' {\em Phys. Lett.} {\bf B736} (2014) 482--487,
\href{http://www.arXiv.org/abs/1404.3542}{{\tt 1404.3542}}.

\bibitem{Hebecker:2014eua}
A.~Hebecker, S.~C. Kraus, and L.~T. Witkowski, ``{D7-Brane Chaotic
  Inflation},'' {\em Phys. Lett.} {\bf B737} (2014) 16--22,
\href{http://www.arXiv.org/abs/1404.3711}{{\tt 1404.3711}}.

\bibitem{Ibanez:2014kia}
L.~E. Ib{\'a}{\~n}ez and I.~Valenzuela, ``{The inflaton as an MSSM Higgs and
  open string modulus monodromy inflation},'' {\em Phys. Lett.} {\bf B736}
  (2014) 226--230,
\href{http://www.arXiv.org/abs/1404.5235}{{\tt 1404.5235}}.

\bibitem{Arends:2014qca}
M.~Arends, A.~Hebecker, K.~Heimpel, S.~C. Kraus, D.~Lust, C.~Mayrhofer,
  C.~Schick, and T.~Weigand, ``{D7-Brane Moduli Space in Axion Monodromy and
  Fluxbrane Inflation},'' {\em Fortsch. Phys.} {\bf 62} (2014) 647--702,
\href{http://www.arXiv.org/abs/1405.0283}{{\tt 1405.0283}}.

\bibitem{McAllister:2014mpa}
L.~McAllister, E.~Silverstein, A.~Westphal, and T.~Wrase, ``{The Powers of
  Monodromy},'' {\em JHEP} {\bf 09} (2014) 123,
\href{http://www.arXiv.org/abs/1405.3652}{{\tt 1405.3652}}.

\bibitem{Franco:2014hsa}
S.~Franco, D.~Galloni, A.~Retolaza, and A.~Uranga, ``{On axion monodromy
  inflation in warped throats},'' {\em JHEP} {\bf 02} (2015) 086,
\href{http://www.arXiv.org/abs/1405.7044}{{\tt 1405.7044}}.

\bibitem{Blumenhagen:2014nba}
R.~Blumenhagen, D.~Herschmann, and E.~Plauschinn, ``{The Challenge of Realizing
  F-term Axion Monodromy Inflation in String Theory},'' {\em JHEP} {\bf 01}
  (2015) 007,
\href{http://www.arXiv.org/abs/1409.7075}{{\tt 1409.7075}}.

\bibitem{Hayashi:2014aua}
H.~Hayashi, R.~Matsuda, and T.~Watari, ``{Issues in Complex Structure Moduli
  Inflation},''
\href{http://www.arXiv.org/abs/1410.7522}{{\tt 1410.7522}}.

\bibitem{Hebecker:2014kva}
A.~Hebecker, P.~Mangat, F.~Rompineve, and L.~T. Witkowski, ``{Tuning and
  Backreaction in F-term Axion Monodromy Inflation},'' {\em Nucl. Phys.} {\bf
  B894} (2015) 456--495,
\href{http://www.arXiv.org/abs/1411.2032}{{\tt 1411.2032}}.

\bibitem{Ibanez:2014swa}
L.~E. Ibanez, F.~Marchesano, and I.~Valenzuela, ``{Higgs-otic Inflation and
  String Theory},'' {\em JHEP} {\bf 01} (2015) 128,
\href{http://www.arXiv.org/abs/1411.5380}{{\tt 1411.5380}}.

\bibitem{Garcia-Etxebarria:2014wla}
I.~Garc{\'\i}a-Etxebarria, T.~W. Grimm, and I.~Valenzuela, ``{Special Points of
  Inflation in Flux Compactifications},'' {\em Nucl. Phys.} {\bf B899} (2015)
  414--443,
\href{http://www.arXiv.org/abs/1412.5537}{{\tt 1412.5537}}.

\bibitem{Blumenhagen:2015kja}
R.~Blumenhagen, A.~Font, M.~Fuchs, D.~Herschmann, E.~Plauschinn, Y.~Sekiguchi,
  and F.~Wolf, ``{A Flux-Scaling Scenario for High-Scale Moduli Stabilization
  in String Theory},'' {\em Nucl. Phys.} {\bf B897} (2015) 500--554,
\href{http://www.arXiv.org/abs/1503.07634}{{\tt 1503.07634}}.

\bibitem{Retolaza:2015sta}
A.~Retolaza, A.~M. Uranga, and A.~Westphal, ``{Bifid Throats for Axion
  Monodromy Inflation},'' {\em JHEP} {\bf 07} (2015) 099,
\href{http://www.arXiv.org/abs/1504.02103}{{\tt 1504.02103}}.

\bibitem{Escobar:2015fda}
D.~Escobar, A.~Landete, F.~Marchesano, and D.~Regalado, ``{Large field
  inflation from D-branes},'' {\em Phys. Rev.} {\bf D93} (2016), no.~8, 081301,
\href{http://www.arXiv.org/abs/1505.07871}{{\tt 1505.07871}}.

\bibitem{Blumenhagen:2015xpa}
R.~Blumenhagen, C.~Damian, A.~Font, D.~Herschmann, and R.~Sun, ``{The
  Flux-Scaling Scenario: De Sitter Uplift and Axion Inflation},'' {\em Fortsch.
  Phys.} {\bf 64} (2016), no.~6-7, 536--550,
\href{http://www.arXiv.org/abs/1510.01522}{{\tt 1510.01522}}.

\bibitem{Escobar:2015ckf}
D.~Escobar, A.~Landete, F.~Marchesano, and D.~Regalado, ``{D6-branes and axion
  monodromy inflation},'' {\em JHEP} {\bf 03} (2016) 113,
\href{http://www.arXiv.org/abs/1511.08820}{{\tt 1511.08820}}.

\bibitem{Hebecker:2015tzo}
A.~Hebecker, J.~Moritz, A.~Westphal, and L.~T. Witkowski, ``{Axion Monodromy
  Inflation with Warped KK-Modes},'' {\em Phys. Lett.} {\bf B754} (2016)
  328--334,
\href{http://www.arXiv.org/abs/1512.04463}{{\tt 1512.04463}}.

\bibitem{Landete:2016cix}
A.~Landete, F.~Marchesano, and C.~Wieck, ``{Challenges for D-brane large-field
  inflation with stabilizer fields},'' {\em JHEP} {\bf 09} (2016) 119,
\href{http://www.arXiv.org/abs/1607.01680}{{\tt 1607.01680}}.

\bibitem{Kaloper:2008fb}
N.~Kaloper and L.~Sorbo, ``{A Natural Framework for Chaotic Inflation},'' {\em
  Phys. Rev. Lett.} {\bf 102} (2009) 121301,
\href{http://www.arXiv.org/abs/0811.1989}{{\tt 0811.1989}}.

\bibitem{Kaloper:2011jz}
N.~Kaloper, A.~Lawrence, and L.~Sorbo, ``{An Ignoble Approach to Large Field
  Inflation},'' {\em JCAP} {\bf 1103} (2011) 023,
\href{http://www.arXiv.org/abs/1101.0026}{{\tt 1101.0026}}.

\bibitem{Dudas:2014pva}
E.~Dudas, ``{Three-form multiplet and Inflation},'' {\em JHEP} {\bf 12} (2014)
  014,
\href{http://www.arXiv.org/abs/1407.5688}{{\tt 1407.5688}}.

\bibitem{Kaloper:2014zba}
N.~Kaloper and A.~Lawrence, ``{Natural chaotic inflation and ultraviolet
  sensitivity},'' {\em Phys. Rev.} {\bf D90} (2014), no.~2, 023506,
\href{http://www.arXiv.org/abs/1404.2912}{{\tt 1404.2912}}.

\bibitem{Kaloper:2016fbr}
N.~Kaloper and A.~Lawrence, ``{A Monodromy from London},''
\href{http://www.arXiv.org/abs/1607.06105}{{\tt 1607.06105}}.

\bibitem{Dvali:2005an}
G.~Dvali, ``{Three-form gauging of axion symmetries and gravity},''
\href{http://www.arXiv.org/abs/hep-th/0507215}{{\tt hep-th/0507215}}.

\bibitem{Dvali:2005zk}
G.~Dvali, ``{A Vacuum accumulation solution to the strong CP problem},'' {\em
  Phys. Rev.} {\bf D74} (2006) 025019,
\href{http://www.arXiv.org/abs/hep-th/0510053}{{\tt hep-th/0510053}}.

\bibitem{Coleman:1980aw}
S.~R. Coleman and F.~De~Luccia, ``{Gravitational Effects on and of Vacuum
  Decay},'' {\em Phys. Rev.} {\bf D21} (1980)
3305.

\bibitem{Bousso:2000xa}
R.~Bousso and J.~Polchinski, ``{Quantization of four form fluxes and dynamical
  neutralization of the cosmological constant},'' {\em JHEP} {\bf 06} (2000)
  006,
\href{http://www.arXiv.org/abs/hep-th/0004134}{{\tt hep-th/0004134}}.

\bibitem{Brown:1988kg}
J.~D. Brown and C.~Teitelboim, ``{Neutralization of the Cosmological Constant
  by Membrane Creation},'' {\em Nucl. Phys.} {\bf B297} (1988)
787--836.

\bibitem{Duncan:1989ug}
M.~J. Duncan and L.~G. Jensen, ``{Four Forms and the Vanishing of the
  Cosmological Constant},'' {\em Nucl. Phys.} {\bf B336} (1990)
100--114.

\bibitem{Feng:2000if}
J.~L. Feng, J.~March-Russell, S.~Sethi, and F.~Wilczek, ``{Saltatory relaxation
  of the cosmological constant},'' {\em Nucl. Phys.} {\bf B602} (2001)
  307--328,
\href{http://www.arXiv.org/abs/hep-th/0005276}{{\tt hep-th/0005276}}.

\bibitem{ArkaniHamed:2006dz}
N.~Arkani-Hamed, L.~Motl, A.~Nicolis, and C.~Vafa, ``{The String landscape,
  black holes and gravity as the weakest force},'' {\em JHEP} {\bf 06} (2007)
  060,
\href{http://www.arXiv.org/abs/hep-th/0601001}{{\tt hep-th/0601001}}.

\bibitem{Ibanez:2015fcv}
L.~E. Ibanez, M.~Montero, A.~Uranga, and I.~Valenzuela, ``{Relaxion Monodromy
  and the Weak Gravity Conjecture},'' {\em JHEP} {\bf 04} (2016) 020,
\href{http://www.arXiv.org/abs/1512.00025}{{\tt 1512.00025}}.

\bibitem{Hebecker:2015zss}
A.~Hebecker, F.~Rompineve, and A.~Westphal, ``{Axion Monodromy and the Weak
  Gravity Conjecture},'' {\em JHEP} {\bf 04} (2016) 157,
\href{http://www.arXiv.org/abs/1512.03768}{{\tt 1512.03768}}.

\bibitem{Brown:2016nqt}
J.~Brown, W.~Cottrell, G.~Shiu, and P.~Soler, ``{Tunneling in Axion
  Monodromy},'' {\em JHEP} {\bf 10} (2016) 025,
\href{http://www.arXiv.org/abs/1607.00037}{{\tt 1607.00037}}.

\bibitem{delaFuente:2014aca}
A.~de~la Fuente, P.~Saraswat, and R.~Sundrum, ``{Natural Inflation and Quantum
  Gravity},'' {\em Phys. Rev. Lett.} {\bf 114} (2015), no.~15, 151303,
\href{http://www.arXiv.org/abs/1412.3457}{{\tt 1412.3457}}.

\bibitem{Rudelius:2014wla}
T.~Rudelius, ``{On the Possibility of Large Axion Moduli Spaces},'' {\em JCAP}
  {\bf 1504} (2015), no.~04, 049,
\href{http://www.arXiv.org/abs/1409.5793}{{\tt 1409.5793}}.

\bibitem{Rudelius:2015xta}
T.~Rudelius, ``{Constraints on Axion Inflation from the Weak Gravity
  Conjecture},'' {\em JCAP} {\bf 1509} (2015), no.~09, 020,
\href{http://www.arXiv.org/abs/1503.00795}{{\tt 1503.00795}}.

\bibitem{Montero:2015ofa}
M.~Montero, A.~M. Uranga, and I.~Valenzuela, ``{Transplanckian axions!?},''
  {\em JHEP} {\bf 08} (2015) 032,
\href{http://www.arXiv.org/abs/1503.03886}{{\tt 1503.03886}}.

\bibitem{Brown:2015iha}
J.~Brown, W.~Cottrell, G.~Shiu, and P.~Soler, ``{Fencing in the Swampland:
  Quantum Gravity Constraints on Large Field Inflation},'' {\em JHEP} {\bf 10}
  (2015) 023,
\href{http://www.arXiv.org/abs/1503.04783}{{\tt 1503.04783}}.

\bibitem{Bachlechner:2015qja}
T.~C. Bachlechner, C.~Long, and L.~McAllister, ``{Planckian Axions and the Weak
  Gravity Conjecture},'' {\em JHEP} {\bf 01} (2016) 091,
\href{http://www.arXiv.org/abs/1503.07853}{{\tt 1503.07853}}.

\bibitem{Hebecker:2015rya}
A.~Hebecker, P.~Mangat, F.~Rompineve, and L.~T. Witkowski, ``{Winding out of
  the Swamp: Evading the Weak Gravity Conjecture with F-term Winding
  Inflation?},'' {\em Phys. Lett.} {\bf B748} (2015) 455--462,
\href{http://www.arXiv.org/abs/1503.07912}{{\tt 1503.07912}}.

\bibitem{Brown:2015lia}
J.~Brown, W.~Cottrell, G.~Shiu, and P.~Soler, ``{On Axionic Field Ranges,
  Loopholes and the Weak Gravity Conjecture},'' {\em JHEP} {\bf 04} (2016) 017,
\href{http://www.arXiv.org/abs/1504.00659}{{\tt 1504.00659}}.

\bibitem{Junghans:2015hba}
D.~Junghans, ``{Large-Field Inflation with Multiple Axions and the Weak Gravity
  Conjecture},'' {\em JHEP} {\bf 02} (2016) 128,
\href{http://www.arXiv.org/abs/1504.03566}{{\tt 1504.03566}}.

\bibitem{Palti:2015xra}
E.~Palti, ``{On Natural Inflation and Moduli Stabilisation in String Theory},''
  {\em JHEP} {\bf 10} (2015) 188,
\href{http://www.arXiv.org/abs/1508.00009}{{\tt 1508.00009}}.

\bibitem{Heidenreich:2015nta}
B.~Heidenreich, M.~Reece, and T.~Rudelius, ``{Sharpening the Weak Gravity
  Conjecture with Dimensional Reduction},'' {\em JHEP} {\bf 02} (2016) 140,
\href{http://www.arXiv.org/abs/1509.06374}{{\tt 1509.06374}}.

\bibitem{Kooner:2015rza}
K.~Kooner, S.~Parameswaran, and I.~Zavala, ``{Warping the Weak Gravity
  Conjecture},'' {\em Phys. Lett.} {\bf B759} (2016) 402--409,
\href{http://www.arXiv.org/abs/1509.07049}{{\tt 1509.07049}}.

\bibitem{Heidenreich:2015wga}
B.~Heidenreich, M.~Reece, and T.~Rudelius, ``{Weak Gravity Strongly Constrains
  Large-Field Axion Inflation},'' {\em JHEP} {\bf 12} (2015) 108,
\href{http://www.arXiv.org/abs/1506.03447}{{\tt 1506.03447}}.

\bibitem{Heidenreich:2016aqi}
B.~Heidenreich, M.~Reece, and T.~Rudelius, ``{Evidence for a Lattice Weak
  Gravity Conjecture},''
\href{http://www.arXiv.org/abs/1606.08437}{{\tt 1606.08437}}.

\bibitem{Montero:2016tif}
M.~Montero, G.~Shiu, and P.~Soler, ``{The Weak Gravity Conjecture in three
  dimensions},''
\href{http://www.arXiv.org/abs/1606.08438}{{\tt 1606.08438}}.

\bibitem{Hebecker:2016dsw}
A.~Hebecker, P.~Mangat, S.~Theisen, and L.~T. Witkowski, ``{Can Gravitational
  Instantons Really Constrain Axion Inflation?},''
\href{http://www.arXiv.org/abs/1607.06814}{{\tt 1607.06814}}.

\bibitem{Saraswat:2016eaz}
P.~Saraswat, ``{The Weak Gravity Conjecture and Effective Field Theory},''
\href{http://www.arXiv.org/abs/1608.06951}{{\tt 1608.06951}}.

\bibitem{Buchmuller:2015oma}
W.~Buchmuller, E.~Dudas, L.~Heurtier, A.~Westphal, C.~Wieck, and M.~W. Winkler,
  ``{Challenges for Large-Field Inflation and Moduli Stabilization},'' {\em
  JHEP} {\bf 04} (2015) 058,
\href{http://www.arXiv.org/abs/1501.05812}{{\tt 1501.05812}}.

\bibitem{Andriot:2015aza}
D.~Andriot, ``{A no-go theorem for monodromy inflation},'' {\em JCAP} {\bf
  1603} (2016) 025,
\href{http://www.arXiv.org/abs/1510.02005}{{\tt 1510.02005}}.

\bibitem{palti}
F.~Baume and E.~Palti, ``{Backreacted Axion Field Ranges in String Theory},''
  {\em JHEP} {\bf 08} (2016) 043,
\href{http://www.arXiv.org/abs/1602.06517}{{\tt 1602.06517}}.

\bibitem{4forms}
S.~Bielleman, L.~E. Ibanez, and I.~Valenzuela, ``{Minkowski 3-forms, Flux
  String Vacua, Axion Stability and Naturalness},'' {\em JHEP} {\bf 12} (2015)
  119,
\href{http://www.arXiv.org/abs/1507.06793}{{\tt 1507.06793}}.

\bibitem{Carta:2016ynn}
F.~Carta, F.~Marchesano, W.~Staessens, and G.~Zoccarato, ``{Open string
  multi-branched and K{\"a}hler potentials},'' {\em JHEP} {\bf 09} (2016) 062,
\href{http://www.arXiv.org/abs/1606.00508}{{\tt 1606.00508}}.

\bibitem{Quevedo:1996uu}
F.~Quevedo and C.~A. Trugenberger, ``{Phases of antisymmetric tensor field
  theories},'' {\em Nucl. Phys.} {\bf B501} (1997) 143--172,
\href{http://www.arXiv.org/abs/hep-th/9604196}{{\tt hep-th/9604196}}.

\bibitem{Garcia-Valdecasas:2016voz}
E.~Garc{\'\i}a-Valdecasas and A.~Uranga, ``{On the 3-form formulation of axion
  potentials from D-brane instantons},''
\href{http://www.arXiv.org/abs/1605.08092}{{\tt 1605.08092}}.

\bibitem{Bielleman:2016grv}
S.~Bielleman, L.~E. Ibanez, F.~G. Pedro, I.~Valenzuela, and C.~Wieck, ``{The
  DBI Action, Higher-derivative Supergravity, and Flattening Inflaton
  Potentials},'' {\em JHEP} {\bf 05} (2016) 095,
\href{http://www.arXiv.org/abs/1602.00699}{{\tt 1602.00699}}.

\bibitem{Buchmuller:2014vda}
W.~Buchmuller, C.~Wieck, and M.~W. Winkler, ``{Supersymmetric Moduli
  Stabilization and High-Scale Inflation},'' {\em Phys. Lett.} {\bf B736}
  (2014) 237--240,
\href{http://www.arXiv.org/abs/1404.2275}{{\tt 1404.2275}}.

\bibitem{Dudas:2015lga}
E.~Dudas and C.~Wieck, ``{Moduli backreaction and supersymmetry breaking in
  string-inspired inflation models},'' {\em JHEP} {\bf 10} (2015) 062,
\href{http://www.arXiv.org/abs/1506.01253}{{\tt 1506.01253}}.

\bibitem{McAllister:2016vzi}
L.~McAllister, P.~Schwaller, G.~Servant, J.~Stout, and A.~Westphal, ``{Runaway
  Relaxion Monodromy},''
\href{http://www.arXiv.org/abs/1610.05320}{{\tt 1610.05320}}.

\bibitem{Blumenhagen:2015qda}
R.~Blumenhagen, A.~Font, M.~Fuchs, D.~Herschmann, and E.~Plauschinn, ``{Towards
  Axionic Starobinsky-like Inflation in String Theory},'' {\em Phys. Lett.}
  {\bf B746} (2015) 217--222,
\href{http://www.arXiv.org/abs/1503.01607}{{\tt 1503.01607}}.

\bibitem{palti2}
D.~Klaewer and E.~Palti, ``{Super-Planckian Spatial Field Variations and
  Quantum Gravity},''
\href{http://www.arXiv.org/abs/1610.00010}{{\tt 1610.00010}}.

\bibitem{Vafa:2005ui}
C.~Vafa, ``{The String landscape and the swampland},''
\href{http://www.arXiv.org/abs/hep-th/0509212}{{\tt hep-th/0509212}}.

\bibitem{Ooguri:2006in}
H.~Ooguri and C.~Vafa, ``{On the Geometry of the String Landscape and the
  Swampland},'' {\em Nucl. Phys.} {\bf B766} (2007) 21--33,
\href{http://www.arXiv.org/abs/hep-th/0605264}{{\tt hep-th/0605264}}.

\bibitem{Gates:1980ay}
S.~J. Gates, Jr., ``{Super p-form Gauge Superfields},'' {\em Nucl. Phys.} {\bf
  B184} (1981)
381--390.

\bibitem{Gates:1980az}
S.~J. Gates, Jr. and W.~Siegel, ``{Variant Superfield Representations},'' {\em
  Nucl. Phys.} {\bf B187} (1981)
389--396.

\bibitem{Deo:1985ix}
B.~B. Deo and S.~J. Gates, ``{Comments on Nonminimal N=1 Scalar Multiplets},''
  {\em Nucl. Phys.} {\bf B254} (1985)
187--200.

\bibitem{Binetruy:1996xw}
P.~Binetruy, F.~Pillon, G.~Girardi, and R.~Grimm, ``{The Three form multiplet
  in supergravity},'' {\em Nucl. Phys.} {\bf B477} (1996) 175--202,
\href{http://www.arXiv.org/abs/hep-th/9603181}{{\tt hep-th/9603181}}.

\bibitem{Ovrut:1997ur}
B.~A. Ovrut and D.~Waldram, ``{Membranes and three form supergravity},'' {\em
  Nucl. Phys.} {\bf B506} (1997) 236--266,
\href{http://www.arXiv.org/abs/hep-th/9704045}{{\tt hep-th/9704045}}.

\bibitem{Binetruy:2000zx}
P.~Binetruy, G.~Girardi, and R.~Grimm, ``{Supergravity couplings: A Geometric
  formulation},'' {\em Phys. Rept.} {\bf 343} (2001) 255--462,
\href{http://www.arXiv.org/abs/hep-th/0005225}{{\tt hep-th/0005225}}.

\bibitem{Cerdeno:2003us}
D.~G. Cerdeno, A.~Knauf, and J.~Louis, ``{A Note on effective N=1
  superYang-Mills theories versus lattice results},'' {\em Eur. Phys. J.} {\bf
  C31} (2003) 415--420,
\href{http://www.arXiv.org/abs/hep-th/0307198}{{\tt hep-th/0307198}}.

\bibitem{Girardi:2007rs}
G.~Girardi, R.~Grimm, B.~Labonne, and J.~Orloff, ``{Correspondence between
  3-form and non-minimal multiplet in supersymmetry},'' {\em Eur. Phys. J.}
  {\bf C55} (2008) 95--99,
\href{http://www.arXiv.org/abs/0712.1923}{{\tt 0712.1923}}.

\bibitem{Nishino:2009zz}
H.~Nishino and S.~Rajpoot, ``{Alternative auxiliary fields for chiral
  multiplets},'' {\em Phys. Rev.} {\bf D80} (2009)
127701.

\bibitem{Bandos:2011fw}
I.~A. Bandos and C.~Meliveo, ``{Three form potential in (special) minimal
  supergravity superspace and supermembrane supercurrent},'' {\em J. Phys.
  Conf. Ser.} {\bf 343} (2012) 012012,
\href{http://www.arXiv.org/abs/1107.3232}{{\tt 1107.3232}}.

\bibitem{Groh:2012tf}
K.~Groh, J.~Louis, and J.~Sommerfeld, ``{Duality and Couplings of
  3-Form-Multiplets in N=1 Supersymmetry},'' {\em JHEP} {\bf 05} (2013) 001,
\href{http://www.arXiv.org/abs/1212.4639}{{\tt 1212.4639}}.

\bibitem{Bachas:2016ffl}
C.~Bachas and T.~Tomaras, ``{Band Structure in Yang-Mills Theories},'' {\em
  JHEP} {\bf 05} (2016) 143,
\href{http://www.arXiv.org/abs/1603.08749}{{\tt 1603.08749}}.

\bibitem{Grimm:2004ua}
T.~W. Grimm and J.~Louis, ``{The Effective action of type IIA Calabi-Yau
  orientifolds},'' {\em Nucl. Phys.} {\bf B718} (2005) 153--202,
\href{http://www.arXiv.org/abs/hep-th/0412277}{{\tt hep-th/0412277}}.

\bibitem{Louis:2002ny}
J.~Louis and A.~Micu, ``{Type 2 theories compactified on Calabi-Yau threefolds
  in the presence of background fluxes},'' {\em Nucl. Phys.} {\bf B635} (2002)
  395--431,
\href{http://www.arXiv.org/abs/hep-th/0202168}{{\tt hep-th/0202168}}.

\bibitem{Villadoro:2005cu}
G.~Villadoro and F.~Zwirner, ``{N=1 effective potential from dual type-IIA
  D6/O6 orientifolds with general fluxes},'' {\em JHEP} {\bf 06} (2005) 047,
\href{http://www.arXiv.org/abs/hep-th/0503169}{{\tt hep-th/0503169}}.

\bibitem{DeWolfe:2005uu}
O.~DeWolfe, A.~Giryavets, S.~Kachru, and W.~Taylor, ``{Type IIA moduli
  stabilization},'' {\em JHEP} {\bf 07} (2005) 066,
\href{http://www.arXiv.org/abs/hep-th/0505160}{{\tt hep-th/0505160}}.

\bibitem{Camara:2005dc}
P.~G. Camara, A.~Font, and L.~E. Ibanez, ``{Fluxes, moduli fixing and MSSM-like
  vacua in a simple IIA orientifold},'' {\em JHEP} {\bf 09} (2005) 013,
\href{http://www.arXiv.org/abs/hep-th/0506066}{{\tt hep-th/0506066}}.

\bibitem{Ibanez:2012zz}
L.~E. Ibanez and A.~M. Uranga, {\em {String theory and particle physics: An
  introduction to string phenomenology}}.
\newblock Cambridge University Press,
2012.
\newblock

\bibitem{Blumenhagen:2016bfp}
R.~Blumenhagen, D.~Herschmann, and F.~Wolf, ``{String Moduli Stabilization at
  the Conifold},'' {\em JHEP} {\bf 08} (2016) 110,
\href{http://www.arXiv.org/abs/1605.06299}{{\tt 1605.06299}}.

\bibitem{Bizet:2016paj}
N.~Cabo~Bizet, O.~Loaiza-Brito, and I.~Zavala, ``{Mirror quintic vacua:
  hierarchies and inflation},'' {\em JHEP} {\bf 10} (2016) 082,
\href{http://www.arXiv.org/abs/1605.03974}{{\tt 1605.03974}}.

\bibitem{Grimm:2008dq}
T.~W. Grimm, T.-W. Ha, A.~Klemm, and D.~Klevers, ``{The D5-brane effective
  action and superpotential in N=1 compactifications},'' {\em Nucl. Phys.} {\bf
  B816} (2009) 139--184,
\href{http://www.arXiv.org/abs/0811.2996}{{\tt 0811.2996}}.

\bibitem{Grimm:2011dx}
T.~W. Grimm and D.~Vieira~Lopes, ``{The N=1 effective actions of D-branes in
  Type IIA and IIB orientifolds},'' {\em Nucl. Phys.} {\bf B855} (2012)
  639--694,
\href{http://www.arXiv.org/abs/1104.2328}{{\tt 1104.2328}}.

\bibitem{Kerstan:2011dy}
M.~Kerstan and T.~Weigand, ``{The Effective action of D6-branes in N=1 type IIA
  orientifolds},'' {\em JHEP} {\bf 06} (2011) 105,
\href{http://www.arXiv.org/abs/1104.2329}{{\tt 1104.2329}}.

\bibitem{Camara:2011jg}
P.~G. Camara, L.~E. Ibanez, and F.~Marchesano, ``{RR photons},'' {\em JHEP}
  {\bf 09} (2011) 110,
\href{http://www.arXiv.org/abs/1106.0060}{{\tt 1106.0060}}.

\bibitem{Marchesano:2014bia}
F.~Marchesano, D.~Regalado, and G.~Zoccarato, ``{U(1) mixing and D-brane linear
  equivalence},'' {\em JHEP} {\bf 08} (2014) 157,
\href{http://www.arXiv.org/abs/1406.2729}{{\tt 1406.2729}}.

\bibitem{Grimm:2010ks}
T.~W. Grimm, ``{The N=1 effective action of F-theory compactifications},'' {\em
  Nucl. Phys.} {\bf B845} (2011) 48--92,
\href{http://www.arXiv.org/abs/1008.4133}{{\tt 1008.4133}}.

\bibitem{Jockers:2004yj}
H.~Jockers and J.~Louis, ``{The Effective action of D7-branes in N = 1
  Calabi-Yau orientifolds},'' {\em Nucl. Phys.} {\bf B705} (2005) 167--211,
\href{http://www.arXiv.org/abs/hep-th/0409098}{{\tt hep-th/0409098}}.

\bibitem{Jockers:2005zy}
H.~Jockers and J.~Louis, ``{D-terms and F-terms from D7-brane fluxes},'' {\em
  Nucl. Phys.} {\bf B718} (2005) 203--246,
\href{http://www.arXiv.org/abs/hep-th/0502059}{{\tt hep-th/0502059}}.

\bibitem{Corvilain:2016kwe}
P.~Corvilain, T.~W. Grimm, and D.~Regalado, ``{Shift-symmetries and gauge
  coupling functions in orientifolds and F-theory},''
\href{http://www.arXiv.org/abs/1607.03897}{{\tt 1607.03897}}.

\bibitem{GarciaEtxebarria:2012zm}
I.~Garcia-Etxebarria, H.~Hayashi, R.~Savelli, and G.~Shiu, ``{On quantum
  corrected Kahler potentials in F-theory},'' {\em JHEP} {\bf 03} (2013) 005,
\href{http://www.arXiv.org/abs/1212.4831}{{\tt 1212.4831}}.

\bibitem{Giddings:2001yu}
S.~B. Giddings, S.~Kachru, and J.~Polchinski, ``{Hierarchies from fluxes in
  string compactifications},'' {\em Phys. Rev.} {\bf D66} (2002) 106006,
\href{http://www.arXiv.org/abs/hep-th/0105097}{{\tt hep-th/0105097}}.

\bibitem{new}
S.~Bielleman, L.~E. Ibanez, F.~G. Pedro, I.~Valenzuela, and C.~Wieck, ``To
  appear,''.

\bibitem{Bielleman:2015lka}
S.~Bielleman, L.~E. Ibanez, F.~G. Pedro, and I.~Valenzuela, ``{Multifield
  Dynamics in Higgs-otic Inflation},'' {\em JHEP} {\bf 01} (2016) 128,
\href{http://www.arXiv.org/abs/1505.00221}{{\tt 1505.00221}}.

\end{thebibliography}\endgroup
\bibliographystyle{utphys}

\end{document}